\documentclass[acmsmall]{acmart}

\AtBeginDocument{%
  \providecommand\BibTeX{{%
    \normalfont B\kern-0.5em{\scshape i\kern-0.25em b}\kern-0.8em\TeX}}}

\setcopyright{acmcopyright}
\copyrightyear{2023}
\acmYear{2023}
\acmDOI{10.1145/1122445.1122456}

\acmJournal{TOSEM}
\acmVolume{X}
\acmNumber{Y}
\acmArticle{1}
\acmMonth{12}

\newcommand{\chen}[1]{\textcolor{red}{#1}}
{\ignorespaces}

\usepackage{algorithm}
\usepackage{algpseudocode}

\algnewcommand\algorithmicforeach{\textbf{for each}}
\algdef{S}[FOR]{ForEach}[1]{\algorithmicforeach\ #1\ \algorithmicdo}
\usepackage{amsmath}
\usepackage{subfigure} 
\usepackage{multirow}
\usepackage{booktabs}
\newcommand{\todo}[1]{}
\renewcommand{\todo}[1]{{\color{red} TODO: {#1}}}

\setcopyright{none}
\renewcommand\footnotetextcopyrightpermission[1]{} 

\begin{document}

\title{Automated Mapping of Adaptive App GUIs from Phones to TVs}

\author{Han Hu}
\email{han.hu@monash.edu}
\affiliation{%
  \institution{Monash University}
  \streetaddress{Wellington Road}
  \city{Clayton}
  \state{Victoria}
  \country{Australia}
  \postcode{3800}
}

\author{Ruiqi Dong}
\email{rdong@swin.edu.au}
\affiliation{%
  \institution{Swinburne University of Technology}
  \streetaddress{John Street}
  \city{Hawthorn}
  \state{Victoria}
  \country{Australia}
  \postcode{3122}
}

\author{John Grundy}
\email{John.Grundy@monash.edu}
\affiliation{%
  \institution{Monash University}
  \streetaddress{Wellington Road}
  \city{Clayton}
  \state{Victoria}
  \country{Australia}
  \postcode{3800}
}

\author{Thai Minh Nguyen}
\email{mngu0072@student.monash.edu}
\affiliation{%
  \institution{Monash University}
  \streetaddress{Wellington Road}
  \city{Clayton}
  \state{Victoria}
  \country{Australia}
  \postcode{3800}
}

\author{Huaxiao Liu}
\email{liuhuaxiao@jlu.edu.cn}
\affiliation{%
  \institution{Jilin University}
  \streetaddress{No.2699, Qianjin Road}
  \city{Changchun}
  \state{Jilin}
  \country{PR China}
  \postcode{130015}
}

\author{Chunyang Chen}
\email{chunyang.chen@monash.edu}
\affiliation{%
  \institution{Monash University}
  \streetaddress{Wellington Road}
  \city{Clayton}
  \state{Victoria}
  \country{Australia}
  \postcode{3800}
}

\renewcommand{\shortauthors}{Hu, et al.}

\begin{abstract}
  With the increasing interconnection of smart devices, users often desire to adopt the same app on quite different devices for identical tasks, such as watching the same movies on both their smartphones and TVs.
  However, the significant differences in screen size, aspect ratio, and interaction styles make it challenging to adapt Graphical User Interfaces (GUIs) across these devices. Although there are millions of apps available on Google Play, only a few thousand are designed to support smart TV displays. Existing techniques to map a mobile app GUI to a TV either adopt a responsive design, which struggles to bridge the substantial gap between phone and TV or use mirror apps for improved video display, which requires hardware support and extra engineering efforts. Instead of developing another app for supporting TVs, we propose a semi-automated approach to generate corresponding adaptive TV GUIs, given the phone GUIs as the input. Based on our empirical study of GUI pairs for TVs and phones in existing apps, we synthesize a list of rules for grouping and classifying phone GUIs, converting them to TV GUIs, and generating dynamic TV layouts and source code for the TV display. Our tool is not only beneficial to developers but also to GUI designers, who can further customize the generated GUIs for their TV app development. An evaluation and user study demonstrate the accuracy of our generated GUIs and the usefulness of our tool.  
\end{abstract}



\maketitle
\section{Introduction}
Graphical User Interfaces (GUI) are ubiquitous in almost all modern desktop software, mobile applications (apps), and online websites.
They provide a visual bridge between a software application and end-users through which they can interact with each other.
A good visual design makes an application attractive and easy to use, which significantly affects the success of the application and the loyalty of its users~\cite{goldschmidt1994visual}.
With the pervasiveness and diversity of devices, especially mobile ones, users tend to use the same app across different platforms in different scenarios (e.g., watching YouTube on a phone on the bus while using it on a tablet or TV at home)~\cite{joorabchi2013real}.
When utilizing apps on TV and tablets, especially those within the media, education, and tools categories, the adaptation to a large-screen TV/tablet display substantially elevates the user experience.
Therefore, to ensure usability,  software GUIs need to be adaptive to various screen sizes and densities, ranging from small 4.7-inch smartwatches to 75-inch smart TVs.

\begin{figure}[!htb]
    \centering
    \includegraphics[width = 0.7\linewidth]{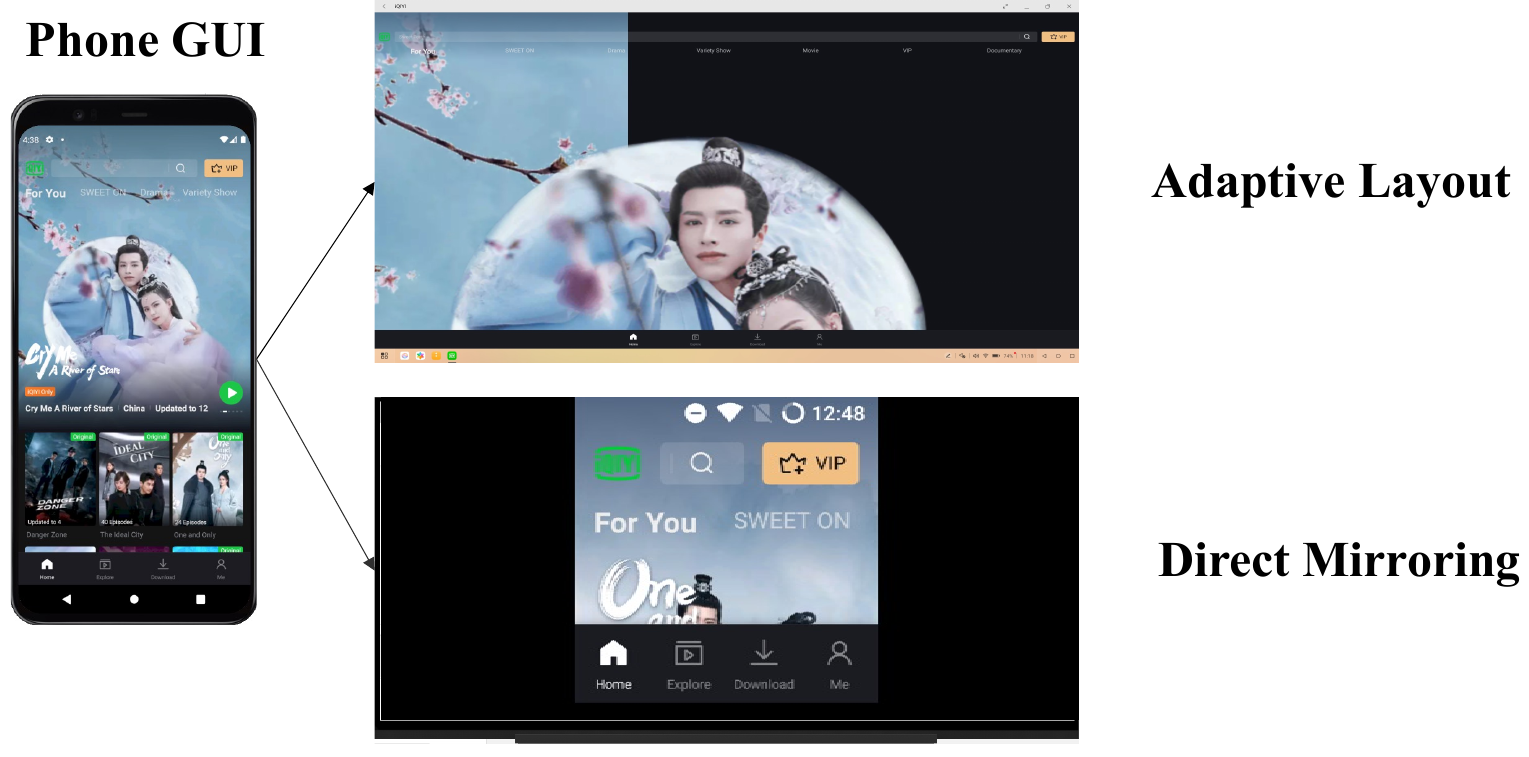}
    \caption{Examples of current large screen GUI conversions}
    \label{fig:currentExamples}
\end{figure}

Although one app may run well on different platforms with the same operating system (e.g., Android), the content may not display well without customization for different devices or screens.
For example, one app designed for the Android phone will only be displayed in the middle of the TV screen with a large dark space on the two sides due to the different aspect ratios as seen in Figure~\ref{fig:currentExamples} (\emph{Direct Mirroring}). 
Additionally, smart TVs and mobile phones interact differently. 
Mobile phones employ fingers to touch and swipe, while smart TVs require remote controls.
TVs need to be redesigned GUIs to accommodate different interactions as well as differences in size and layout.

Currently, there are three ways commonly used to deal with this issue.
First, some GUI development frameworks provide support of responsive and adaptive design~\cite{adaptiveUI} for developers to adapt the GUI to any possible screen size, such as Android's Material Design~\cite{laine2021responsive} and iOS Auto Layout~\cite{iosAdaptive}.
The \emph{Adaptive Layout} in Figure~\ref{fig:currentExamples} shows an example of the adaptive layout.
While these techniques adjust the GUI according to device screen size, they can inadvertently magnify or misalign UI components, diminishing the user experience. Moreover, as TVs employ interaction modalities distinct from mobile devices, like remote controls or voice commands, such designs may lack requisite support. 
This solution, primarily tailored for devices with minor screen size variations, such as mobile phones and tablets, may not be ideal for TVs, thereby presenting challenges in mobile-to-TV app adaptation.


Second, certain developers choose to devise platform-distinct GUIs or craft unique apps tailored to the TV interface. 
The task of originating a dedicated TV app and the associated upkeep is resource-intensive and time-consuming.
Considering the commonalities in functionalities between mobile and TV apps and the potential to leverage extant designs and resources, there exists a strong rationale for repurposing assets, potentially yielding significant developmental cost savings.


Third, there are some screen mirroring apps (e.g., Chromecast~\cite{chromecast}), which are used to cast music, video, play games, and display photos on a big screen.
However, these are effective primarily for video projection rather than general app screens.
Phone manufacturers like Samsung, Apple, Huawei, and OnePlus have introduced the concept of Desktop Mode~\cite{huaweiDesk, Dex} for cross-screen GUI conversion via their phones.
Desktop Mode maps the current phone GUIs to larger external screens via an HDMI adaptor or WiFi by redesigning GUI layouts individually to improve the user experience with a larger screen.
The essence of these mirroring technologies revolves around either the pre-configuration of layouts for specific apps or the employment of adaptive layout techniques for projection.

To overcome these limitations, we propose an approach for generating the appropriate GUI layout for Android TV based on the existing GUI design for phones.
Unlike responsive design, where a screen 'flows' from a phone design into a larger device, adaptive projection in our approach offers tailor-made solutions. 
Given the app GUI of a page on a mobile phone, our approach automatically generates the corresponding GUI code to make it adaptive to the TV screen.
The development team, including the visual designers and developers, will benefit from our approach.
Once they finish the phone app development, our approach can automatically generate the GUI design and TV implementation for any phone page.
Designers and developers can easily customize it to enhance their productivity without starting from scratch.

To build the tool, we first investigate how many apps currently support smart TV in 5,580 popular apps from Google Play\cite{googleplay}.
We find only 5.34\% of apps support running on smart TV.
Second, we perform a formative study on the characteristics of current TV-phone apps and GUIs.
We collect 1405 TV-phone app pairs from Google Play and Dangbei\cite{dangbei} but only collect 589 TV-phone GUI pairs with clear GUI correspondence in these app pairs.
We then summarize 12 and 9 categories of GUI components groups common on mobile and TV, respectively, and analyzed their corresponding conversion.
Based on the empirical study results, we build our tool in the following steps: group isolated GUI components parsed from GUI metadata captured by UI Automator~\cite{uiautomator}, convert phone GUI groups into corresponding TV groups, optimize their layouts by OR constraints formulas~\cite{jiang2019orc}, and translate current TV GUI to our language - and platform free GUI domain-specific language (DSL) for further development. 

We choose two of the best-known and most commonly used technologies: desktop mode and direct mapping as the baselines for the evaluation.
We select mIoU, overall satisfaction, and structure rationality as the metrics to evaluate the converted TV GUIs by our approach and baselines.
We also recruited 20 participants with extensive experience in Android GUI development, design, and use to evaluate the converted TV GUIs. 
The experimental results show that TV GUIs converted by our approach achieve 21.05\%, 10.42\%, and 21.31\% improvement in mIoU, overall satisfaction, and structure rationality than the best current conversion techniques.
Besides, a pilot user study also provides the initial evidence of the usefulness of our tool for bootstrapping adaptive GUI from phone to TV.

Our contributions to this work are summarized as follows:
\begin{itemize}    
    \item As far as we know, this is the first study on automated GUI conversion from smartphone to TV GUI; 
    
    \item We propose a new approach to generate TV-hosted GUIs from Android mobile phone GUIs;
    
    \item We carry out an empirical study to understand the current status of GUI support in the phone app to TV display and how the GUI mapping patterns between two platforms; and
    
    \item We demonstrate the effectiveness of our approach with extensive automated evaluation and manual checking. We also provide initial evidence of our tool's usefulness via a pilot user study.
\end{itemize}



\section{Empirical study of GUIs between phones and TVs}
\label{sec:empiricalStudy}
To better understand the current status and characteristics of GUI adaptation to smart TV, we conduct a large-scale empirical study to answer two questions: 1) How many phone apps support TV display? 2) How do app GUIs change between TVs and smartphones?

\subsection{RQ1 How many phone apps support TV displays?}
\label{sec:supportTV}

To answer RQ1, we analyze a large number of industrial Android apps from all 33 categories on Google Play~\cite{googleplay}.
We crawl the top 200 most popular apps in each category (at the time of Jun 2021).
Since some apps are not free, we obtain 5,580 apps that support Android phones by default.

According to the official guidelines of developing Android TV~\cite{AndroidTVDe}, it is compulsory for apps running on TV to declare a TV activity with an intent filter \textit{CATEGORY\_LEANBACK\_LAUNCHER} in the Android manifest file of Android projects.
In addition to the declaration, TV-enabled apps often have separate TV layouts called \textit{layout-television} and \textit{layout-tv}.
Therefore, we decompile Android APKs to check whether the specified intent filters or layout XML files exist.
We find only 298 of these apps supporting TV display, accounting for only 5.34\% of the total.
These 298 apps belong to 29 categories including \textit{Weather} (16.78\%), \textit{Education} (15.10\%), \textit{Tool} (11.07\%).
 
Even if an app advertises that it supports a TV display, this does not imply that it is appropriately optimized for the TV.
We manually check 298 apps that claim to support TV displays.
We physically run the apps on a smart TV and critically evaluate how well various GUI pages adapt to the TV display. GUI pages exhibiting discrepancies such as substantial black margins at the screen's bilateral extents, discordant aspect ratio components, disordered layouts, and unoptimized navigation, among other factors -- which all may diminish the user experience on TV displays -- are regarded as inadequate instances of the TV adaptation.
We discover only 11 of 298 apps allow TV display on all GUI pages.
287 out of 298 apps modify a few representative GUI pages, such as the home or landing page, to accommodate TV displays.
These 287 apps have an average of 22.3 Android activities, but we only find support for TV display in an average of 4 Android activities.
Other GUIs of these apps also look poor on TV, with large black margins on both screen sides, and mismatched aspect ratio components, as the \emph{Direct mapping} in Figure~\ref{fig:currentExamples}.
Due to the significant difference between phone and TV displays, some development teams tend to develop separate apps for different platforms.

\begin{figure}[!htbp]
    \centering
    \includegraphics[width = 0.9\linewidth]{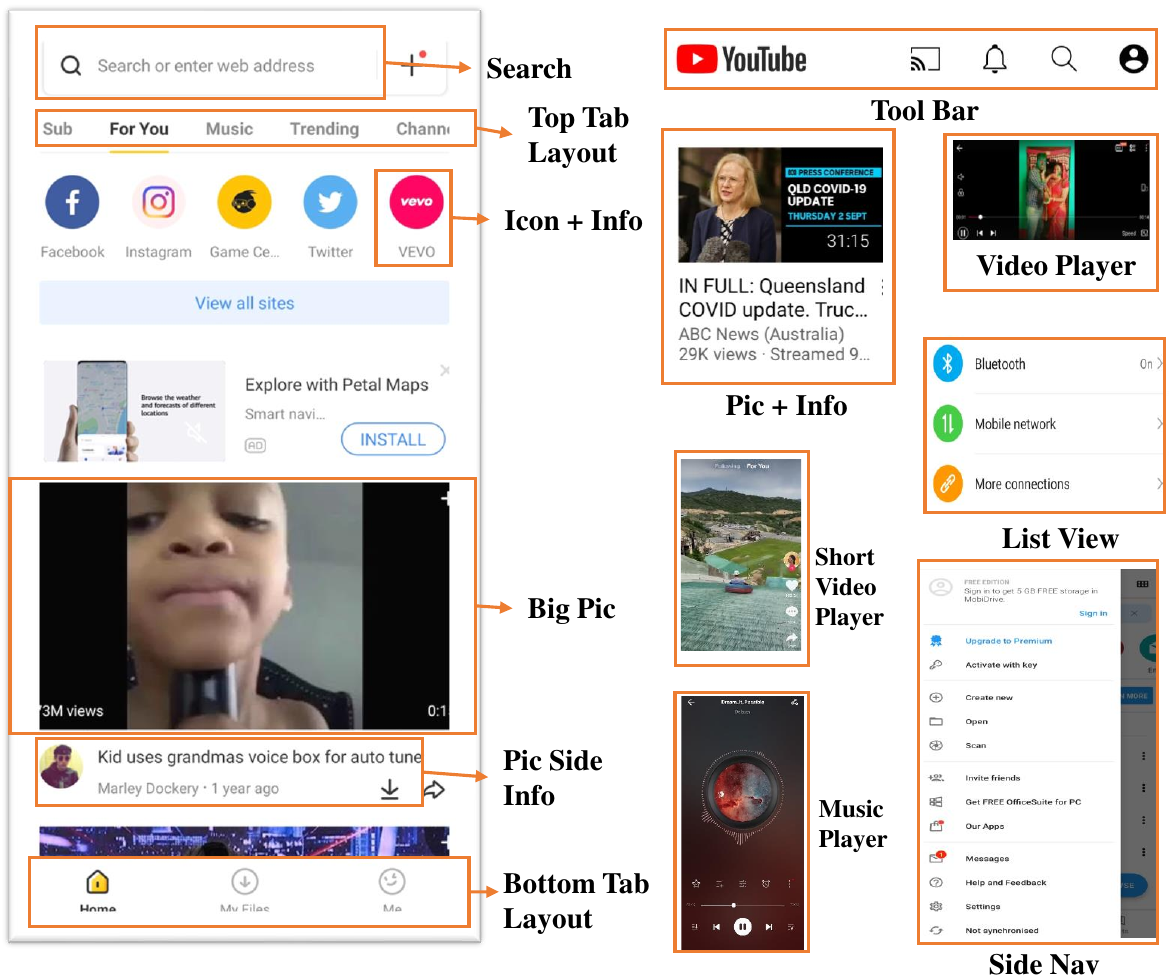}
    \caption{Examples of Phone GUI Groups}
    \label{fig:phoneGroups}
\end{figure}

\subsection{RQ2 How do app GUIs change between TVs and smartphones?}
\label{sec:rq2}
We collected custom-designed TV apps and their matching phone apps for a comparative study since there are relatively few apps that support both TV and phone.

\subsubsection{TV-phone App Pairs Collection}
\label{sec: appCollection}
We collected 249 TV apps from Google Play's TV category~\cite{googleTV} and 2,556 from Dangbei~\cite{dangbei}, which is one of the largest TV app stores. We eliminated any apps that have not been updated for over two years.

We then  matched the TV apps' corresponding smartphone apps on Google Play.
To begin, we search for phone apps with the same app name, developers, and category.
Second, if the corresponding phone version app cannot be found, we broaden our search to include apps with similar app names developed by the same developers.
Finally, for the TV apps that do not find the phone-version apps, four volunteers manually collect their matching phone apps.

We match 1,405 TV-phone app pairs, with the three most common categories being \emph{Video (42\%)}, \emph{Education (23\%)} and \emph{Tool (21\%)}.
Video apps, such as \textit{Youtube}, \textit{IQIYI}, are most suitable for smart TVs due to the characteristics of TV itself, so video apps have become the most popular apps on TV.
Because smart TVs are mostly used at home, there are a number of educational apps for children.
Tool apps, like TV app store, remote control and projection control are also ubiquitous on TVs.

\subsubsection{TV-phone GUI Pair Collection}
\label{sec:guiCollection}
We use the DroitBot~\cite{li2017droidbot}, Fastbot~\cite{cai2020fastbot} and Uiautomator2~\cite{uiautomator2} to automatically explore apps and collect rendered screenshots and metadata of GUIs in apps.
The metadata is a documentary object model (DOM) tree of current GUIs, which includes the hierarchy and properties (e.g., class, bounding box, layout) of UI components. 
We can infer the GUI hierarchy from the DOM Tree hierarchy in metadata.
After removing duplicates, we obtain 6,697 Android GUI data and 4,112 TV GUI data.
We notice that most TV apps simplify or restructure their GUIs to accommodate various usage scenarios and requirements of smart TVs during this process.

At TV-phone GUI pairing, we exploit the semantic similarity of Android activity names to pair the TV and phone GUIs automatically.
Then, we automatically compare UI components on GUI pages in order to match state-level GUI pairings in a lesser granularity.
Finally, we manually check the automatically discovered pairs and select the final valid TV-phone pairs.
We extract activity names from each GUI and encode them into numerical semantic vectors using a pre-trained BERT~\cite{devlin2019bert} model.
Then, we match the TV-phone GUI pairs by comparing their semantically close activity vectors.
For example, the GUI in activity \textit{homeActivity} and \textit{mainActivity} are matched by close semantic vectors.
However, one Android activity may have multiple Android fragments~\cite{fragment} and GUI states~\cite{machiry2013dynodroid, GUIstate} with different UI components and layouts in current industrial apps.
We further compare GUI components between phones and TVs to pair TV-phone GUIs at lower granularity.
In pairing, UI components are identified by their types and properties.
UI components between phones and TVs  with the same types and properties are considered paired GUI components.
For example, two \emph{TextViews} with the same texts, two \emph{ImageViews} with the same images, two \emph{Buttons} with the same texts are considered the paired components.
If more than half of the UI components in two GUIs are paired, they are considered a state-level TV-phone GUI pair.
Finally, we manually checked all discovered pairs and identified 589 TV-phone state-level GUI pairs with clear GUI correspondence between phone and TV components.

\subsubsection{GUI component grouping}
\label{sec:GUIgroup}
A series of UI components that are near in position and hierarchy of the GUI and tend to have the same functionality are referred to as GUI groups\footnote{Sometimes called GUI patterns}~\cite{guiBuild, neil2014mobile}.
Exploring these group changes based on UI groups rather than each individual component is more beneficial~\cite{neil2014mobile}.
So, we must first identify and categorize GUI groups that are common to both TV and phone before summarising the guidelines for GUI changes from phone to TV.

\begin{figure}[!htbp]
    \centering
    \includegraphics[width = 0.9\linewidth, height = 7cm]{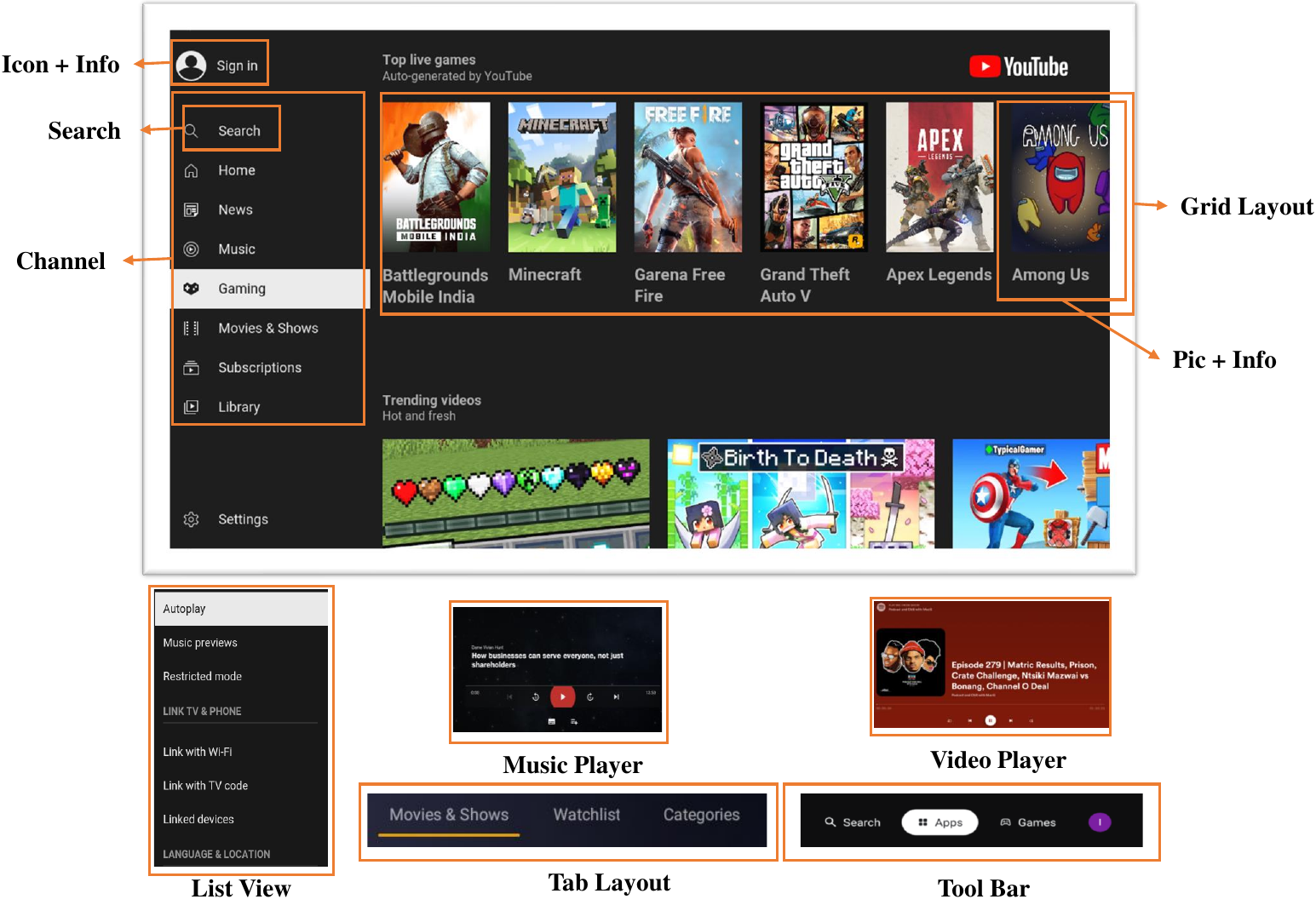}
    \caption{Examples of Key TV GUI Groups}
    \label{fig:TVGroups}
\end{figure}

We perform an iterative-open coding process, which is widely used to generate categories in Software Engineering~\cite{seaman1999qualitative, hu2023first, hu2023look, chen2021my}. We do this on 120 randomly selected phone GUIs and 80 TV GUIs (approximately 2\% of collected phone and TV GUIs) to categorize their GUI groups.
Four volunteers with Android design experience undertook three steps in our open coding procedure.
At first, we are inspired by Google design~\cite{GoogleMaterial} and development guidelines~\cite{androidGuidelines}, every volunteer categorizes GUI groups in selected GUIs individually.
After the initial coding, let four volunteers have a discussion and merge conflicts.
They clarify scope boundaries among categories and misunderstandings in this step.
In the third step, they iterate to revise classifications and discuss with each other until a consensus is reached.
Finally, we determined 12 phone group types: \textit{Icon + Info}, \textit{Tool Bar},  \textit{Bottom Tab Layout}, \textit{Search}, \textit{Top Tab Layout}, \textit{Pic Side Info}, \textit{Pic + Info}, \textit{Side Nav}, \textit{Short Video Player}, \textit{Video/Music Player}, \textit{Big Pic} and \textit{List View} and 9 TV group types: \textit{Icon + Info}, \textit{Tool Bar}, \textit{Search}, \textit{Tab Layout}, \textit{Channel}, \textit{Grid Layout}, \textit{Pic + Info}, \textit{Video/Music Player} and \textit{List View}.
Figure~\ref{fig:phoneGroups} and \ref{fig:TVGroups} show examples of summarized phone and TV groups.
These phone and TV GUI group categories can be divided into two subcategories.
GUI groups in the first subcategories are widely used in both phone and TV, e.g. \textit{Tool Bar}, \textit{Search}, and \textit{Video/Music Player}.
GUI groups in the second subcategory only exist in phones or TVs respectively, e.g. such as \textit{Bottom Tab Layout} and \textit{Channel}.

To verify the accuracy of GUI grouping and classification, we first randomly sample 20 TV and smartphone apps in collected 1405 TV-phone app pairs in Section~\ref{sec: appCollection}, respectively.
The selected apps have covered all 6 TV app categories (Tool, Video, Music \& Audio, Education, Entertainment, and Weather) listed on Google Play~\cite{ggTV}.
Then, we manually calculate the distribution of each GUI group in selected TV and smartphone apps.
Finally, we find that these summarized GUI components groups have covered 93.69\% and 93.47\% of GUI groups on phones and TVs, respectively.
In our analysis, we observe the presence of uniquely shaped GUI components in both phone and TV interfaces. These components resist typical categorization due to their complex functional requirements in production and everyday use. They constitute 6.31\% and 6.53\% of phone and TV GUIs, respectively. These specific GUI components primarily appear within individual apps or among apps from the same developer. We categorize these components under an $Others$ group category for the purpose of our study. To address these edge cases in the GUI conversion process, we propose the development of default templates. Given the relatively minor percentage of GUI components in the $Others$ category and their limited appearance in specific applications, we assert that our categorization effectively represents the vast majority of commonly used phone and TV GUIs.

Table~\ref{tab:matchRules} shows the details of GUI group distributions.
Subcolumns \emph{Group} and \emph{Distribution} of columns \emph{Phone} and \emph{TV} denote the GUI group categories and distribution in the experiment.
TV's subcolumn \emph{Group} does not contain the \emph{Others} GUI group, which comprises 6.53\% of all TV GUIs in the experiment.
On phones, the most popular categories of components groups are \textit{Icon + Info} (13.31\%), \textit{Tool Bar} (11.41\%) and \textit{List View} (11.14\%), but on TV, categories \textit{Pic + Info} (19.13\%), \textit{Grid Layout} (13.37\%) and \textit{List View} (13.18\%) are most common categories.
The official guideline of TV GUI design~\cite{AndroidTVDe} suggests two principles for TV GUI design \emph{All TV GUIs should display in landscape mode} and
\emph{The core TV GUIs use card-like views instead of ListView or ViewPager to make better use of horizontal screen space and accommodate TV interaction}.
Standard current TV GUIs obey these two principles to use more grid layouts and card-like widgets.
Thus card-like categories \textit{Pic + Info} and \textit{Grid Layout} are more popular on TVs than on phones.


\setlength{\textfloatsep}{10pt}

\subsubsection{Component group alignment}
\label{sec:groupAlignment}
After conducting our GUI components grouping study, we notice that the design principles of TV and phone GUIs are vastly different, resulting in no obvious one-to-one alignments between most TV-phone GUI groups.
Furthermore, the contents of one phone GUI group may be dispersed throughout numerous groups in the TV GUI, and vice versa.
Therefore, we summarize heuristic rules from our collected TV-phone GUI pairs for automatic GUI mapping from phone to TV.

Firstly, we randomly divide the 589 TV-phone GUI pairs into experimental and validation sets in an 8:2 ratio.
In the experimental set, four volunteers follow the same three steps to perform open coding to analyze and extract conversion rules.
Table~\ref{tab:matchRules} shows the extracted GUI group match rules from phone to TV.
To accommodate how the TV and remote interact, each TV GUI group uses card-like views.
The converted TV GUI group recalculates the new size depending on the quantity and types of components in the existing GUI to fit the TV screen size. 
According to the GUI group study in Section~\ref{sec:GUIgroup}, we use \textit{Grid Layout} as our default template for mapping.
Component groups with the same meaning \textit{Icon + Info}, \textit{Tool Bar}, \textit{Search}, \textit{Top Tab Layout}, \textit{Video/Music Player} and \textit{List View} in phone and TV are transferred directly.
According to the characteristics and the official design guideline~\cite{AndroidTVDe} of TV, \textit{Pic Side Info}, \textit{Pic + Info} and \textit{Big Pic} are all converted to \textit{Pic+Info} in TV.
After exploration, current TV GUIs tend to replace components in the phone's \textit{Side Navigation} and \textit{Bottom Tab Layout} to \textit{Channel} in TV,
so we follow this trend.
\textit{Short Video Player} should use customized templates in TV, but there's no such TV app with this GUI feature at the moment.
As a result, we don't provide the corresponding TV group individually at the moment, instead relying on \textit{Video Player} to convert.

We use the validation set to verify these mapping rules.
Our first step involves manually identifying and extracting GUI components from respective groups within the phone GUI. 
Subsequently, we locate corresponding components in the matching TV GUI. 
Our final process entails verifying if these TV GUI components align with the anticipated TV groups and comply with the matching rules established in our experimental set. 
Given \emph{m} instances of \emph{Side Nav} groups in our validation set, and $n$ corresponding TV GUI groups classified as $Channel$ groups, we compute the mapping rule accuracy as the ratio $n/m$.
Note that if the phone GUI group is eliminated in the corresponding TV GUI, the case is considered invalid and will not be counted.
The \emph{Mapping Accuracy} in Table~\ref{tab:matchRules} demonstrates the correctness rate of each mapping rule.
Finally, we find that the correctness of rules 1, 2, 3, 4, 5 and 7 are 96\%, 99\%, 91\%, 99\%, 99\%, 99\% and 100\%, respectively, indicating that these direct mapping rules are accurate and universal.
Rules 8, 9, 10, 11, and 12 have an accuracy of 83\%, 95\%, and 87\%, 90\%, and 99\%, respectively, suggesting these change rules are also accurate and common.

\begin{table}[!htbp]
\setlength{\abovecaptionskip}{0pt}
\setlength{\belowcaptionskip}{0pt}
\caption{Component group matching between phone and TV. TV's subcolumn \emph{Group} does not contain the \emph{Others} GUI group, which comprises 6.53\% of all TV GUIs in the experiment.}
\scalebox{0.8}{
\begin{tabular}{|c|cc|cc|c|}
\toprule
\multirow{2}{*}{\textbf{Index}} & \multicolumn{2}{c|}{\textbf{Phone}}     & \multicolumn{2}{c|}{\textbf{TV}}  & \multirow{2}{*}{\textbf{Mapping Accuracy}}                     \\
~ & Group & Distribution & Group & Distribution & \\

\midrule
1 & Icon + Info  & 13.31\%       & Icon + Info  & 8.32\%  & 96\%                   \\
2 & Tool Bar  & 11.41\%          & Tool Bar    & 7.83\%   & 99\%                   \\
3 & List View      & 11.14\%     & List View          & 13.18\%  & 91\%             \\
4 & Top Tab Layout  & 8.88\%    & Top Tab Layout   & 7.68\%  & 99\%               \\
5 & Search    & 7.98\%          & Search       & 7.12\%  & 99\%                   \\
6 & Others & 6.31\% & Grid Layout (Default) & 13.37\% & 90\%  \\
7 & Video/Music Player & 3.50\% & Video/Music Player    & 7.56\%  & 100\%          \\

\midrule
8 & Pic Side Info  & 8.90\%     & \multirow{3}{*}{Pic + Info} & \multirow{3}{*}{19.13\%}  & 83\%   \\
9 & Pic + Info   & 8.67\% &             &         & 95\%             \\
10 & Big Pic     & 3.52\%         &        &      & 87\%                     \\

\midrule
11 & Bottom Tab Layout   & 10.31\% & \multirow{2}{*}{Channel}    & \multirow{2}{*}{9.28\%}   & 90\%   \\
12 & Side Nav  & 6.07\%  &             &        & 99\%             \\
\bottomrule
\end{tabular}%
}
\label{tab:matchRules}
\end{table}

\subsection{Summary and Implications}
\label{sec:implication}
Our empirical study shows that:
(1) Only 5.34\% of popular phone apps support TV displays. 
(2) In TV-phone GUI pairs, there is not much explicit one-to-one correspondence between phone and TV component groups.
(3) We summarize 12 and 9 categories of GUI components groups on phone and TV, covering 93.69\% and 93.47\% popular phone and TV GUIs, respectively. 
(4) We extract and evaluate 12 existing GUI group-mapping rules from phone to TV based on summarised GUI component groups.

The lack of TV-display support for phone apps confirms the necessity of tool development for semi-automated GUI mapping between phone and TV.
That motivates our study and the empirical findings of component group mapping are the backbone of our proposed approach.

\section{Semi-automated TV GUI Generation}
\begin{figure*}
    \centering
    \includegraphics[width = \textwidth]{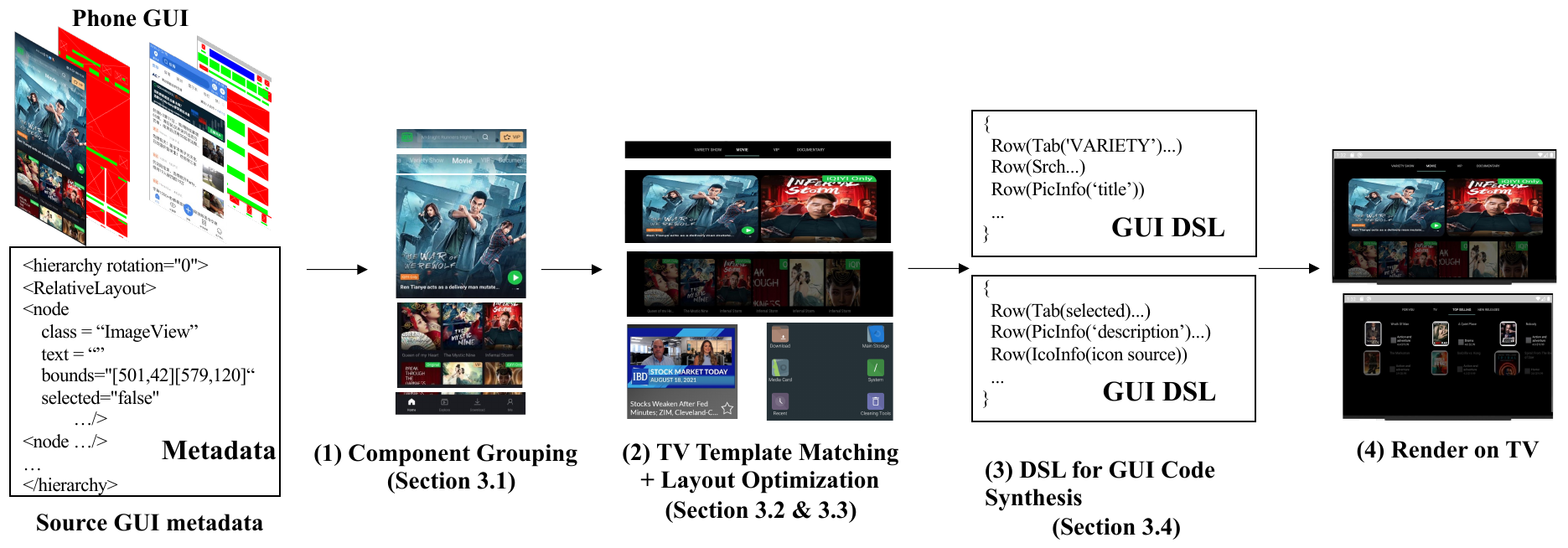}
    \caption{Overview of Automated GUI Conversion from Phone to TV}
    \label{fig:pipeline}
\end{figure*}

Motivated by our empirical findings and the official layout criteria for Android TV~\cite{AndroidTV}, 
we here propose our semi-automated Android-based TV GUI generation approach.
Our approach develops a lightweight migration system that converts run-time GUIs in a series of phases, including component recognition and grouping, template matching, layout optimization, and GUI domain-specific languages (DSL) generation.
The overall pipeline of our approach is illustrated in Figure~\ref{fig:pipeline}.

We first gather the run-time GUI metadata generated by UI Automator\cite{uiautomator}, including screenshots and GUI metadata, and analyze hierarchical metadata to identify GUI components.
Secondly, to construct a suitable GUI mapping, we design a component grouping algorithm that leverages the GUI information extracted from metadata to group isolated elements properly.
Thirdly, we compare elements' attributes and hierarchical similarities to classify grouped elements into appropriate types and match corresponding TV templates.
Then, we optimize the whole GUI layouts to adapt TV screens by OR-constraints~\cite{jiang2019orc, jiang2020orcsolver}.
Finally, given the incompatibility of phone and TV systems, a cross-platform TV GUI DSL is designed to describe generated TV GUI for further compilation and rendering. 

\subsection{Component Grouping}
\label{sec:group}
\begin{figure}
    \centering
    \includegraphics[width = 0.9\linewidth]{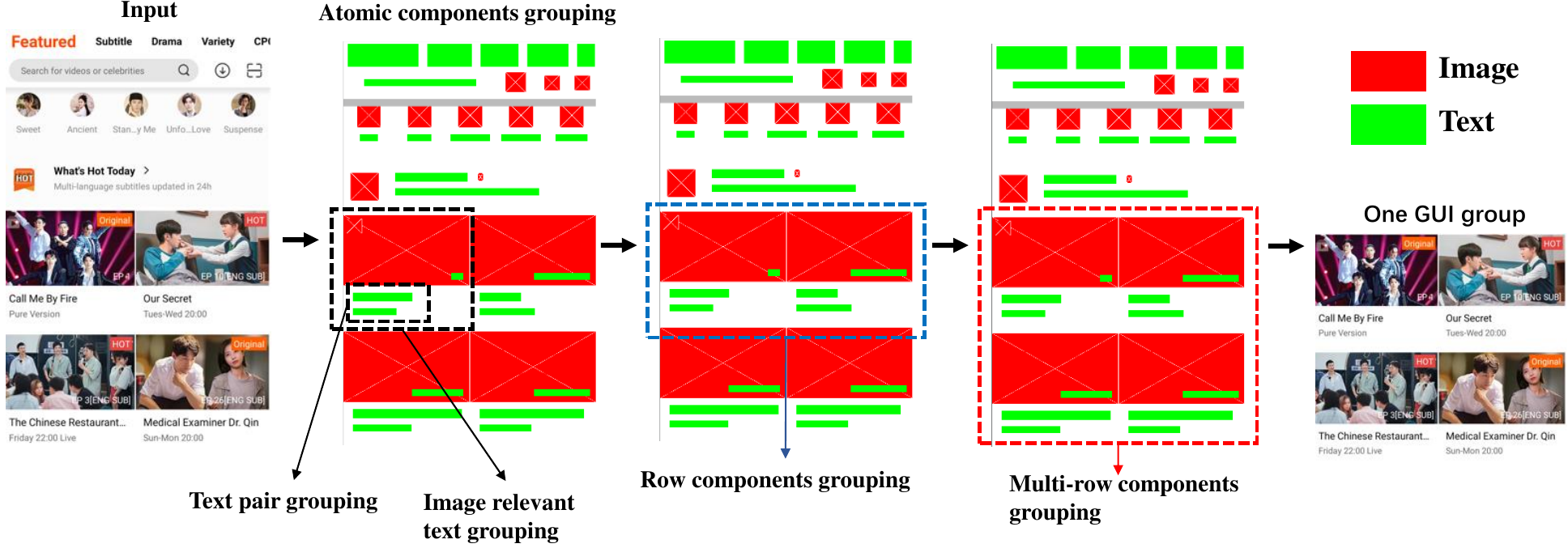}
    \caption{A running example of components grouping. The block dotted box 1 represents atomic grouping, the blue 2 represents row grouping, and the red 3 represents multi-row grouping.
    }
    \label{fig:grouping}
\end{figure}

As mentioned in Section~\ref{sec:GUIgroup}, we first group isolated GUI components to GUI groups as the basic unit of follow-up work.
We parse metadata captured by UI Automator to accurately infer the pix-based coordinate of GUI components' bounding box and classify these components to proper types like \textit{TextView}, \textit{ImageView}, and \textit{Button}.
Once each GUI component's bounding boxes and type in a rendered screen are confirmed, the next phase is to group atomic components with similar domain-specific functions to one component group.
Figure~\ref{fig:grouping} illustrates a running example for our component grouping algorithm. 
Overall, the algorithm consists of three-level grouping with different assembly granularities. 
To begin, we gather all atomic components that have significant relationships, as well as all images and their text descriptions.
Second, we then group components in the same row by their hierarchy, type, and total width.
Third, we then merge adjacent rows with the same hierarchy and pix-based areas.

\subsubsection{Atomic Components Grouping}
Regarding text pairs are very close on the Y axis, they tend to be a pair of descriptions, with the caption at the top and the additional explanation of the caption at the bottom~\cite{neil2014mobile, andriodUI, zhang2021screen}.
Texts below or on the right of images always have strong semantics with them, such as previewed movie names, image descriptions, etc~\cite{neil2014mobile, andriodUI}.
Thus, heuristics are designed to aggregate relevant components based on component type, position, size, and structural relationship.

For text pair grouping, given upper text $T_u$ and below texts set $T_s$, if texts in $T_s$ meet the following requirements, they will be grouped with $T_u$: 
(1) On X-axis, $T_u$ and $T_s$ overlap, or the gap between them must be less than $0.025 \times screen\ horizontal\ resolution$.
(2) On Y-axis, $T_s$ is the closet element around the $T_u$. The gap between them must be lower than $0.025 \times screen\ vertical\ resolution$.
We employ the subsequent three phases to confirm that our heuristics contains sufficient relevant component possibilities and to acquire the best applicable empirical parameters.
Following the relevant work of Xiaoyi etc.~\cite{zhang2021screen} and the official Android GUI design guidelines~\cite{AndroidTV, adaptiveUI}, we first determine all potential positions and distance range of the images and their accompanying isolated UIs.
Second, we randomly selected 10 GUI pages in each category of our collection of phone apps (90 GUI pages in total).
We experiment on the selected data to verify whether our heuristic rule can cover all the relevance possibilities.
The experimental results indicate that our heuristic principles are capable of covering all possibilities in existing GUIs. 
Finally, we assessed the impact of all empirical coefficients within the range provided by Android official guidelines on the selected pages in increments of 0.001 $\times$ the current screen's horizontal and vertical resolution. 
We ultimately determined 0.025 to be the optimal empirical coefficient.
Potential compatibility issues with phones and apps may lead to distorted or misaligned UI components in the GUI, making our heuristic principles inapplicable.
However, these circumstances can be rectified by developers through subsequent development.

For image relevant texts grouping, given texts set $T$ and image $I$, if texts in $T$ meet the following requirements, they will be grouped with $I$:
(1) On Y-axis, $T$ is the closet element below the $I$ and the gap between $T$ and $I$ must be less than $0.025 \times screen\ vertical\ resolution$.
(3) On X-axis, the midpoints of $T$ and $I$ are the same. If not, their overlap on X-axis must be more than 50\% of the smaller one in $T$ and $I$.
(4) If $T$ is in a grouped text pair, pack both texts as the relevant texts of $I$.  

\begin{algorithm}[]
\setlength{\parskip}{0.2cm plus4mm minus3mm}
\caption{Row Components Grouping}
\label{alg:domBased}
\begin{algorithmic}[1]
\Require DOM tree $Tree$, required group width $W_{group}$
\Ensure Row components group set $G$
\State $Leaves \leftarrow []$
\ForEach{component $c \in C$}
    \If{length(getChildren ($c$)) == 0}
        \State $Leaves$.add($c$) \textcolor{gray}{//collect all leaves in $tree$}
    \EndIf
\EndFor
\ForEach {Leaf $l \in Leaves$}
    \State $node_{cur} \leftarrow l$, $W_{cur} \leftarrow $ $l$.width
    \While{$W_{cur} <= W_{group}$}
        \State $p \leftarrow$ getParent($l$) \textcolor{gray}{//Backtrack to the parent node $p$ of $l$ in $Tree$}
        \State $node_{cur} \leftarrow p$, $W_{cur} \leftarrow $ $p$.width
    \EndWhile
    \State $S \leftarrow$ trimTree($node_{cur}$) \textcolor{gray}{//Trim $tree$ at node $node_{cur}$ to get a subtree $S$}
    \State $g_l \leftarrow []$  \textcolor{gray}{// the set to collect all leaves in $S$ }
    \ForEach{$l^{\prime}  \in S$}
    \If{$l^{\prime} \in L$}
    \State $g_l$.add($l^{\prime}$)
    \EndIf
    \EndFor
    \State $G$.add($g_l$ ) \textcolor{gray}{//Add $g_l$ to $G$}
\EndFor
\State \Return $G$
\end{algorithmic}
\end{algorithm}

\subsubsection{Row Components Grouping}

According to Google's Android design guide~\cite{androidGuidelines, andriodUI}, the standard UI design typically treats each row as the fundamental unit. 
Therefore, the UI components in the phone are arranged in rows.
The metadata of the phone's GUI, as illustrated in Figure~\ref{fig:pipeline}, describes its GUI hierarchy in the form of an XML DOM tree.
UI components in the same row are in the same subtree in the DOM tree.
The purpose of our row components grouping is thus to trim the subtrees from the DOM tree that contains all UI components in one row.
When the algorithm finds that components inside the same DOM subtree have occupied an entire row, it will put all of these components into one group.

\begin{figure}
    \centering
    \includegraphics[width = 0.9\linewidth]{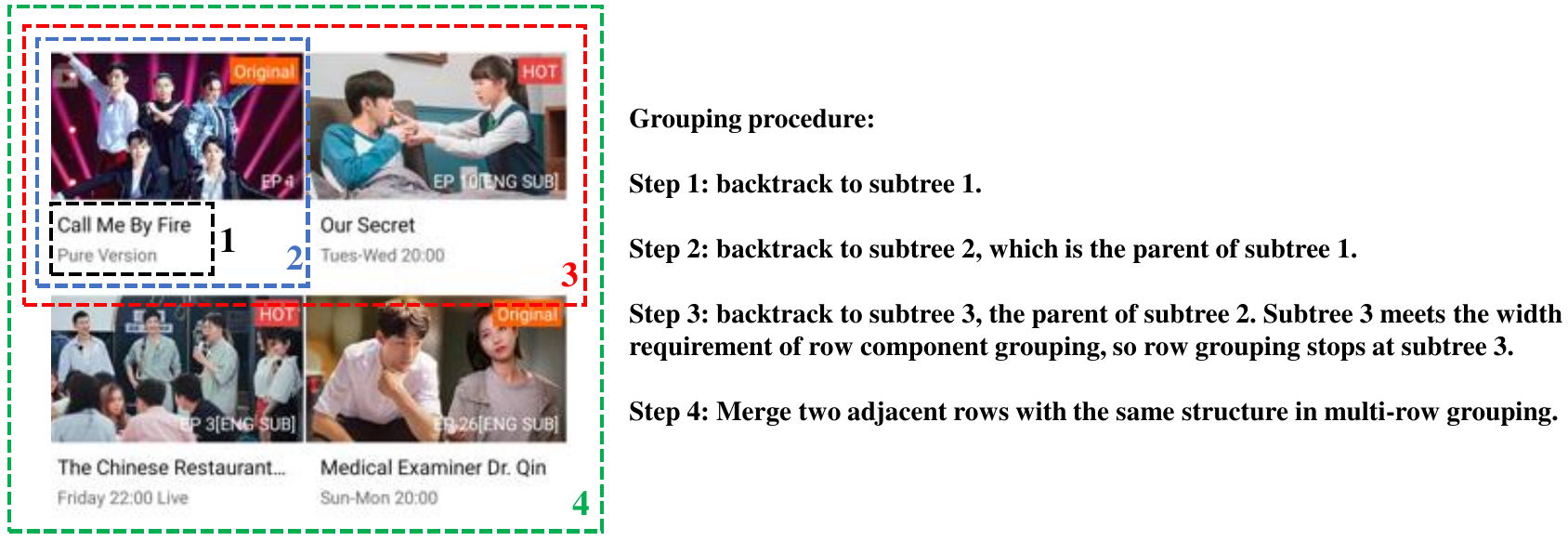}
    \caption{A running example of Algorithm~\ref{alg:domBased} and multi-row component grouping.}
    \label{fig:rowExample}
\end{figure}

\subsection{TV Template Matching}
Algorithm~\ref{alg:domBased} demonstrates our row components grouping process.
It utilizes a DOM tree ($tree$) of metadata and a pre-defined group width parameter ($W_{group}$) as inputs. 
The latter is a critical determinant dictating when to partition a subtree representing a row on the screen. 
Specifically, when the subtree's width extends to match the phone screen's width, it is deemed to occupy an entire row. 
Given that GUI elements are always displayed with margins on both the left and right sides of the screen, we empirically define the requisite width as 0.85 times the total screen width in this work.
The output of Algorithm~\ref{alg:domBased} is the grouped components \emph{G}.
The leaf nodes in the DOM tree are the specific UI components, so Alg.~\ref{alg:domBased} first walks from the root node of the DOM tree to the terminal leaves, collecting all leaves $L$ (lines 1-6).
Following the identification of leaf nodes, the algorithm backtracks up the DOM tree from each leaf in $Leaves$ to establish respective subtrees representative of a screen row.
For each leaf $l$, if the width of the current node ($W_{cur}$) exceeds the predefined group width $W_{group}$, prune at node $l$.
If not, backtrack to the current node's parent node $p$.
If the width of $p$ still does not exceed the predefined group width $W_{group}$, the traceback procedure continues until the current node's width meets the criterion that is larger than the predefined width (lines 7-12).
We then trim the DOM tree from the node where the backtracking stops to get the cropped subtree $S$ (line 13).
To group all UI components within a single row, the algorithm subsequently collects all leaves in the subtree $S$ as a group $g_l$ (lines 14-19).

Figure~\ref{fig:rowExample} shows a running example of the Alg.~\ref{alg:domBased}.
The backtracking starts at leaf TextView \emph{Call Me By Fire} and reaches the leaf's parent subtree 1.
However, the width of subtree 1 is less than the required group width $W_{group}$.
We then keep going back to subtree 2, but the width of subtree 2 is still not satisfied.
All the way back to subtree 3, its width is more than the required group width.
The algorithm thus stops at subtree 3 and trims subtree 3 from the GUI DOM tree as a GUI component row.

\subsubsection{Multi-row Components Grouping}
Components in one component group may be spread across multiple rows, such as \textit{ListView} and \textit{GridLayout}.
In these component groups, the components of different rows tend to have the same structure and component types.
Thus, given two adjacent rows $r_i$ and $r_j$, based on their bounding boxes, we compute the relative positions of the upper left and lower left corners of components in each row. If components in $r_i$ and $r_j$ have the same types and relative upper left and lower left corners, two rows will be grouped.
As shown in Figure~\ref{fig:rowExample}, subtree 3 has the same structure as the row below, so merge these two rows into one GUI group in step 4.
Note that we allow two rows in one group in the implementation with one different component.

We use GUI built-in and visual features to categorize groups and match related templates after component grouping.
We first summarize the unique attributes of each type of group.
The built-in attributes of the group \textit{Top Tab Layout}, \textit{Bottom Tab Layout}, \emph{Tool Bar}, \emph{Search}, \emph{Video/Music Player}, \emph{List View}, and \emph{Side Nav} are their component types, position and relationship with their sibling components. The unique built-in attributes of the groups \emph{Top Tab Layout}, \emph{Bottom Tab Layout} are the component types and their positions in the GUI pages. 
Both groups will use the GUI components of the tab layout class and will be situated in the upper or lower half of the GUI pages, accordingly. 
Similarly, the unique built-in attributes of the groups \emph{Tool Bar}, \emph{Search}, and \emph{Side Nav} are also their specific component types and positions. 
The group \emph{Tool Bar} is located at the top of the GUI page. 
The \emph{Search} group is located in the upper half of the GUI pages with a search box. 
The \emph{Side Nav} group is located on the leftmost side of the GUI page and has a unique side navigation property in the DOM tree. 
The group \emph{Video/Music Player} and  \emph{List View} also have player and list view attributes unique to the Dom tree, respectively. 
 The image size and related information position features are used to classify groups \textit{Icon + Info}, \textit{Pic Side Info}, \textit{Pic + Info}, and \textit{Big Pic}.
 
When matching templates, we count the number of built-in attributes that each group has in common with each template.
After matching all templates, if the maximum number of matching attributes exceeds the threshold, the most comparable template is assigned to the group. 
If it is below the threshold, the group is deemed unrecognized. In the case of unrecognized GUI groups, we provide a general Grid Layout-based template for their conversion.
The threshold is set at 2 after multiple iterations of error correction based on experimental feedback.

\begin{figure}
    \centering
    \includegraphics[width = 0.6\linewidth]{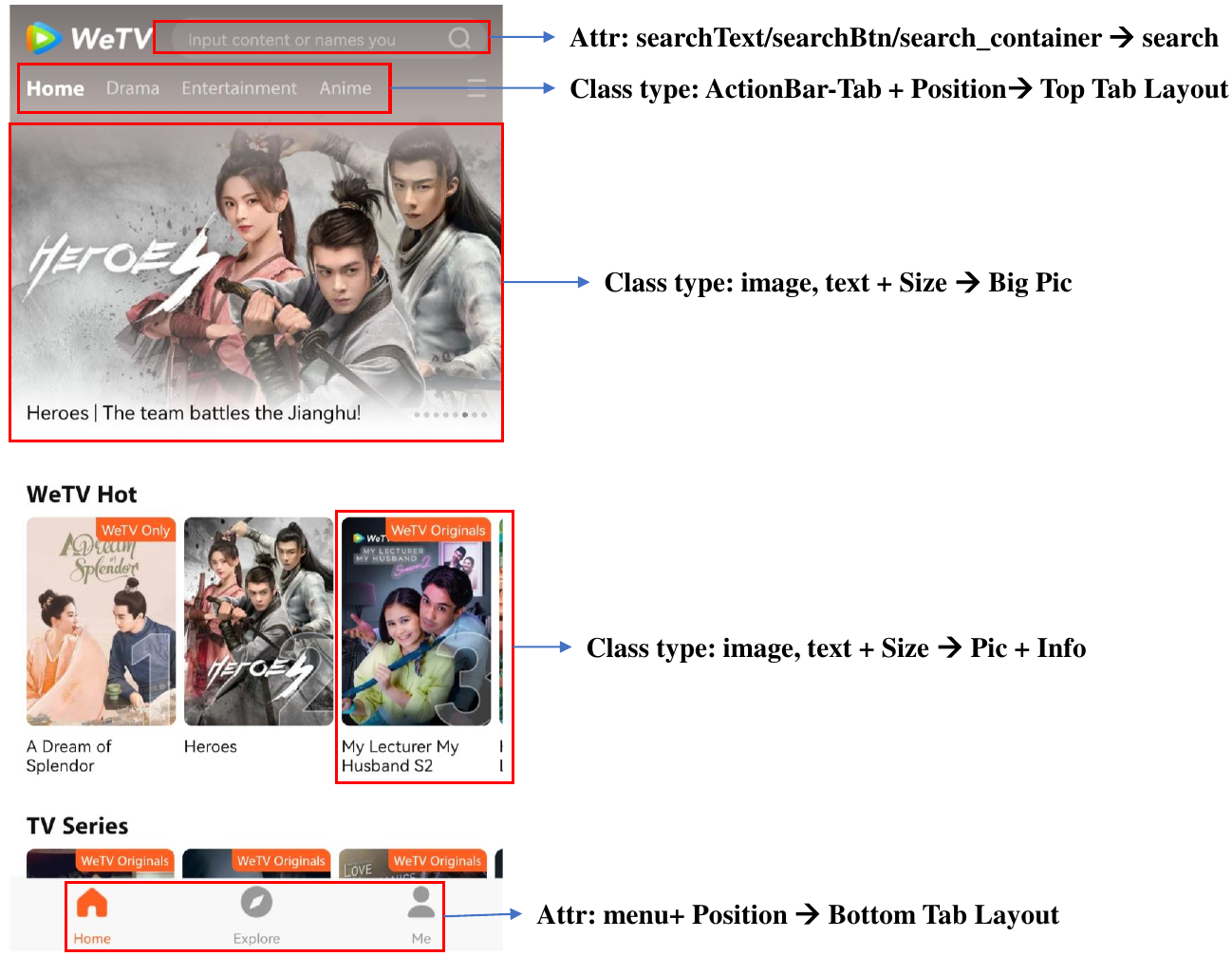}
    \caption{An example of template match for the phone GUI. }
    \label{fig:templateMatch}
\end{figure}

Figure~\ref{fig:templateMatch} demonstrates an example of classifying GUI groups.
We notice when parsing the metadata of this GUI page that there are some GUI components at the top with the fields \emph{searchText}, \emph{searchBtn}, and \emph{search\_container} in their attributes. 
These are similar to the attributes of the \emph{Search} template.
As a result, this GUI group is categorized under the category \emph{Search}.
The top of the page has a GUI group with the class type \emph{ActionBar-Tab}.
Its features of position and class type meet the template \emph{Top Tab Layout}, and hence it is classified to category \emph{Top Tab Layout}.
To proceed, we can now continue to identify GUI groups of categories \emph{Big Pic}, \emph{Pic+Info}, and \emph{Bottom Tab Layout} from the page.

Following the identification of GUI groups in the phone page, we convert these categorized phone GUI groups into matching TV GUI groups using the mapping rule in Table~\ref{tab:matchRules}.

\subsection{Layout optimization}
Due to various screen sizes and design principles, direct layout mapping may lead to issues like large white space on the right or too much content squeezed into a small area.
The Android TV design guidelines~\cite{TVDesign, AndroidTVDe} state that TV layouts should be landscape and have more card-like components since this is more suited to TV interaction and enables the display of only the most essential image and text contents.
So, if we just map each GUI group into the TV screen one by one, or simply change the GUI orientation from portrait to landscape without optimizing the layout, the final produced TV layout would be quite inflexible and violate the design standards of TV GUIs. 

On phone GUIs, huge images frequently take up a whole row, as in Figures~\ref{fig:phoneGroups} and \ref{fig:templateMatch}.
However, the design guidelines for Android TV~\cite{AndroidTV, TVDesign, AndroidTVDe} state that images taking up an entire row would seem to be quite odd and significantly worsen the user experience.
Additionally, if one image is too big, the remaining components will be too small for viewers watching the TV from far away to see.
Currently, as \emph{Desktop mode} in Figure~\ref{fig:currentExamples}, Android's adaptive technology immediately projects onto the TV based on the proportion of its components. 
The slider picture occupied the whole TV screen and entirely obscured the location of other components after being adjusted by the adaptive layout, resulting in an odd overall effect.

To overcome these issues, we propose a TV-based GUI layout OR-constraints~\cite{jiang2019orc, jiang2020orcsolver} to optimize converted TV layouts.
The OR-constraint is a combination of soft constraints, with the whole thing being a hard constraint.
The hard restrictions must be met, but the soft constraints are not required.
We set soft constraints for each component in one row, and all components in one row must meet the hard constraints for the Android TV layout. 
Unlike template-based approaches, which necessitate pre-designing templates and manually specifying rules for when each alternative should be invoked, constraints-based layout optimization can be more flexible and adaptive for a variety of screens without fixed templates and rules.
Different from the Android adaptive GUI layout, the constraint-based layout optimization approach, in conjunction with the TV group template and guidelines, can arrange the layout of UI components on the whole screen. 
This prevents the scenario when some UI components, due to the original size of the phone, are too huge to be transferred to the TV.


We summarize the layout requirements from the TV design guidelines~\cite{andriodUI, AndroidTV, TVDesign} and our empirical study.
We convert these requirements into constraints, and force converted GUI groups to optimize their layouts to satisfy these constraints.
We lift three predefined heuristics as basic constraints for TV GUI layout:
(1) Arrange GUI widgets from left to right.
(2) If there is not enough space in the current row for the following widget, it will begin on the leftmost side of the next row.
(3) Each component should be put within a predefined size range.
Every TV GUI design must adhere to these three hard constraints.
At the same time, TV GUIs also have one soft constraint:
(4) There should be no black gaps on either side of the TV screen, which means the UI Component should fill the whole width of the screen.
If there isn't enough space in the converted TV GUI, components will be removed in order of their size in the phone GUI, from small to large.
The Z3 SMT solver~\cite{de2008z3} is used to solve OR-constrains.

The following part shows how we formulate these constraints.
Given a row with $n$ components $W$.
For each component $w_i \in W$, assume its left $w^{left}_i$, width $w^{width}_i$, top $w^{top}_i$, and current available width in the row $r_a$.
Let $w^{height}_j \in W$ denotes the max components height in the current row.
The maximum/minimum widths and heights of components $w_i$ are represented as $width^{max}_i$, $width^{min}_i$, $height^{max}_i$, and $height^{min}_i$.
For the constraint (1), we convert it to the following formula:
\begin{equation}
\begin{aligned}
    C_{1} := (w^{left}_i = w^{left}_{i-1} + w^{width}_{i-1}) \land (w^{top}_i = w^{top}_{i-1}) 
    \\ \land (r_a >= width^{min}_i)
\end{aligned}
\end{equation}
For the constraint (2), it is formulated as
\begin{equation}
\begin{aligned}
    C_{2} := (w^{left}_i = 0) \land ( w^{top}_i >=  w^{top}_j + w^{height}_j)
    \\
    \land (r_a < width^{min}_i)
\end{aligned}
\end{equation}
For constraint (3), we assign preferred widths, and heights for common GUI component types. 
So,
\begin{equation}
\begin{aligned}
        C_{3} := (width^{min}_i <= w^{width}_i <= width^{max}_i) \land \\(height^{min}_i <= w^{height}_i <= height^{max}_i)
\end{aligned}
\end{equation}

For constraint (4), let $r_{tv}$ represent the maximum width of the TV screen in one row, and its logical expression is
\begin{equation}
    C_{soft}^1 := \sum_{i=1}^n (w_i^{width}) = r_{tv}
\end{equation}

Each row represents a single formula unit. If a GUI group covers multiple rows, the GUI group is referred to as a formula unit.
For each formula unit, $C_1$ and $C_2$ are OR constraints, and $C_3$ is a hard constraint.
In addition to three basic constraints, some weighted soft constraints $C_{soft}$ are followed. 

The final formula is thus:
\begin{equation}
\label{eq:or}
    C_{unit} := ((C_{1}\ \lor \ C_{2})\ \land \ C_{3}) \lor
    C_{soft}^k
\end{equation}
where $k$ represents the number of soft constraints.
We can dynamically apply new soft constraints to Equation~\ref{eq:or} dependent on the demands of the TV GUI.



\subsection{DSL for GUI Code Synthesis}

In the dynamic world of app development, distinct environments and languages often shape apps for phones and TVs. To bridge this gap, we introduce a platform-agnostic GUI Domain Specific Language (DSL) designed for efficient code synthesis~\cite{mernik2005and, hudak1997domain}. This card-style DSL, bolstered by a specialized GUI block library, captures the nuances of TV GUIs. It seamlessly interprets the type, size, position, and relationships of each GUI row, translating each component into a DSL statement. By focusing on the core types and layouts of the GUI components, we ensure the DSL remains streamlined and concise. A standout feature of our DSL is its ability to consistently represent GUI layouts, promoting a unified design language across platforms.

GUI groups, distinguished by their attributes such as position, kind, size, and hierarchical relationships, form the foundation of our DSL representation. To rigorously define our DSL, we use a context-free grammar (CFG) formalism~\cite{engelfriet1997context}.
Within the CFG, non-terminal symbols act as abstract placeholders, encapsulating the essential building blocks of our DSL. The accompanying production rules, on the other hand, precisely dictate how these non-terminals can be transformed or expanded, thereby translating abstract concepts into concrete representations within the GUI layout domain.

\begin{table}[!htp]
\setlength{\abovecaptionskip}{0pt}
\setlength{\belowcaptionskip}{0pt}
\centering
\caption{Context-Free Grammar for our DSL}
\label{tab:cfg_dsl}
\begin{tabular}{|l|l|}
\hline
\textbf{Non-terminal \& Description} & \textbf{Production} \\
\hline
\( S \): Statement structure in DSL & \( S \to L_i (C_1(P_1, P_2, \ldots P_m), \ldots, C_j(P_1, P_2, \ldots, P_m)) \) \\
\hline
\( L \): Layout type & \( L \to \text{Row} | \text{Col} \) \\
\hline
\( C \): GUI group category & $C \to$ ToolBar|List|Tab|Srch|Grid|Player|PicInfo|IcoInfo|Chan \\
\hline
\( P \): Properties of a GUI group\footnotemark & \( P \to \text{Image/Icon\ Title} | \text{Size} | \text{Text} | \text{Selected} | \text{Image/Icon\ Source} \)\\
\hline
\end{tabular}
\end{table}
\footnotetext{The property type of a GUI group is fixed, but its content can vary.}

Table~\ref{tab:cfg_dsl} encapsulates the CFG definition of our DSL.
The column titled \textit{Non-terminal\ \&\ Description} enumerates our grammar's primary constructs alongside concise descriptions. These non-terminals serve as the essential scaffolds in constructing GUI layouts within our DSL. For instance, the non-terminal \textit{S} articulates the DSL's overview statement structure, encapsulating the complete GUI layout. 
\textit{L} designates the layout pattern of a series of GUI groups, distinguishing between horizontal (\textit{Row}) and vertical (\textit{Col}) UI component alignments. 
The category of a TV GUI group is captured by \textit{C}, with categories spanning from \textit{ToolBar} to \textit{Chan}, elucidated in Section~\ref{sec:GUIgroup}. 
Lastly, \textit{P} represents properties associated with \textit{C}, such as the image size or image title in the \textit{Pic+Info} GUI group or the selected state of \textit{TabLayout}.
The GUI group's property type remains constant, yet the precise content within this property exhibits variability. As an illustration, distinct images may possess divergent image titles.
The \textit{Production} column elucidates the transformation rules for each non-terminal, guiding their expansion to depict specific GUI layout elements in the DSL.

\begin{figure}[!htp]
    \centering
    \includegraphics[width = 0.8\linewidth]{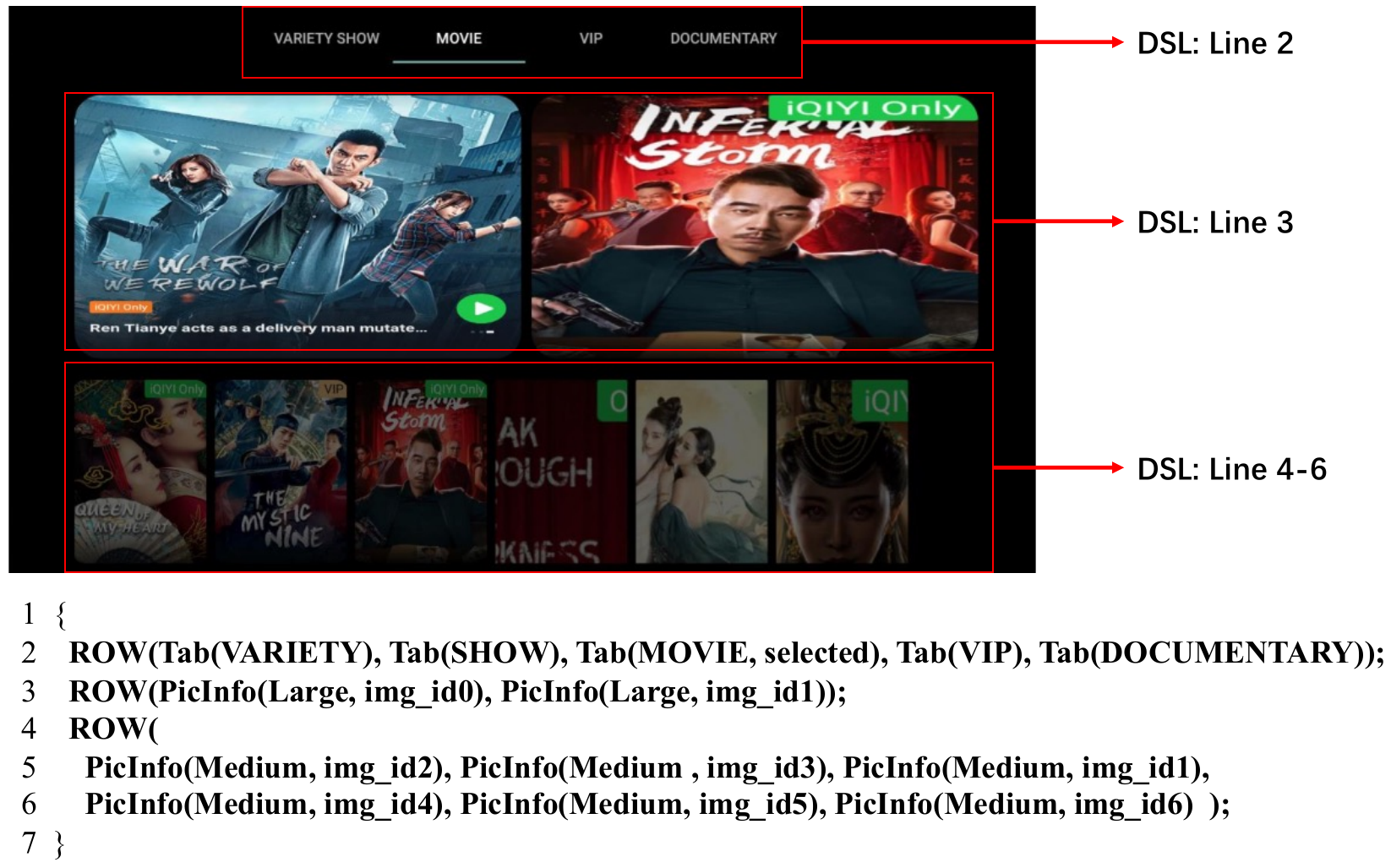}
    \caption{An Example of Our TV GUI DSL}
    \label{fig:dsl}
\end{figure}

Figure~\ref{fig:dsl} exhibits an instance of our lightweight DSL. The second line of DSL encapsulates the layout of the first row in the TV GUI. The term \textit{Row} is our layout phrase ($L$), stipulating the horizontal positioning of all input components. The $Tab$ in $Tab(VARIETY)$ symbolizes the \emph{Top Tab Layout} category in TV GUI groups ($C$), and $VARIETY$ in $Tab(VARIETY)$ signifies the text property ($P$) within the group. 
In lines 3, 4, and 5, $PicInfo$ signifies the category encompassing $Pic + Info$. Within this classification, the image size is determined by the primary parameter, with potential classifications being large, medium, or small. Concurrently, the secondary parameter delineates the image's source.

\setlength{\textfloatsep}{10pt}

The DSLs will be translated into real-world code according to the platform's requirements and render the TV interface.
We pre-write the TV GUI style library and install a client app with the TV GUI style library on TV. 
According to the DSL keywords, the style library calls the associated GUI code, converts it to real-world code, and invokes the system to render it.

\subsection{Implementation}
Uiautomator2\cite{uiautomator2}, Fastbot~\cite{cai2020fastbot}, and Droidbot~\cite{li2017droidbot} are used to capture GUI metadata.
All algorithms in the pipeline are implemented by Python3.
Our pipeline generates DSL for TV layout, then sends the intermediates to TV.
We pre-install a client app on TV to receive generated DSL and translate DSL to real-world code to render TV screens.
The TV client app is built on the Leanback~\cite{leanback} library.
Leanback is a library for TV user interface and is provided by Google.
To produce a GUI, Leanback must be developed based on the interface. With our DSL, Leanback input is automatically generated without the need for the developer to again spend time creating the GUI.
According to the optimized size of each GUI component, each GUI group type is split into three subclasses during implementation: large, medium, and small. 
Based on Leanback, we developed templates for all the subclasses of the 9 TV GUI groups we summarized.
According to the GUI DSL, the client app calls the corresponding template, generates the corresponding TV GUI code, and renders it.








\section{Accuracy Evaluation}
\label{sec:autoEvaluation}

We carry out experiments to evaluate the accuracy of the GUI grouping and the converted TV GUI effects in our pipeline.

\subsection{Accuracy of GUI Grouping}
\label{sec:groupEva}
The accuracy of GUI component grouping serves as the foundation for succeeding techniques.
Hence, we first evaluate the accuracy of our proposed grouping algorithm.

\subsubsection{Procedure}
To confirm the generalizability of our grouping algorithm, we randomly select another 10 apps that are not in our dataset.
To ensure the quality of selected apps, we only select apps with at least one million Google Play installations.
A total of 100 GUI pages are selected as the test dataset, with 10 GUI pages from each app being randomly selected.
Next, we perform the grouping algorithm on the selected GUI pages and collect grouping results. 
The grouping results are then manually checked to ensure their reasonableness. 

\subsubsection{Metric}
A fair result for a GUI group must contain only UI elements with strong logical connections.
When manually verifying the rationality of grouping, a group fails if it contains UI components that are logically unrelated to other UI components inside the group.
When we refer to `logically related UI components', we mean a logical association of GUI elements based on interactivity and role. 
Interactivity pertains to the direct communicative relationship between components, such as an input field and a 'submit' button. The 'role' encapsulates the components' collective function within the broader UI context. Hence, `logically related UI groups' signify clusters of UI elements assembled based on their interaction synergy and shared role in the user interface.

We use the exact match rate to illustrate the percentage of reasonable GUI groups.
The exact match is a binary metric, with a value of 1 if correct and 0 otherwise.
If there are $m$ GUI groups on this GUI page, and $n$ of them are reasonable, then the grouping accuracy of this GUI page is $n/m$.
Figure~\ref{fig:exampleGroup} demonstrates an example of correct and wrong GUI grouping on one GUI page.
Groups 1, 3, and 4 in Figure~\ref{fig:exampleGroup} are considered to be correctly grouped since all UI components within the group are logically related.
However, group 2 is considered to be grouped incorrectly. 
This is because, in addition to the search-related UI components in group 2 (the UI components boxed by the blue dashed line in Figure~\ref{fig:exampleGroup}), there are two logically unrelated icons, which should not be classified into this search-related UI group.
The grouping accuracy of Figure~\ref{fig:exampleGroup} is 0.75.
Considering the subjective nature of determining 'logically related UI components', we employ three individuals with a minimum of one year's experience in GUI development to independently conduct the exact match evaluation. 
Thereafter, the Fleiss Kappa value~\cite{fleiss1971measuring} is used to measure the level of agreement among these three evaluators.
Fleiss Kappa values are interpreted as follows: [0.01, 0.20] signifies slight agreement, (0.20, 0.40] indicates fair agreement, (0.40, 0.60] represents moderate agreement, (0.60, 0.80] suggests substantial agreement, and (0.8, 1] signifies near perfect agreement. 
Instances where the Fleiss Kappa value falls below 0.8 prompt a discussion, analysis, and re-evaluation by the three evaluators until a Fleiss Kappa value exceeding 0.8 is achieved.
To evaluate our method's ability to minimize the isolated UI components, we follow related works~\cite{zhang2021screen, chen2019code} to select the proportion of reduced UI components as the second metric.
Suppose there are $J$ UI components on the original GUI pages, and $K$ UI components and UI groups are left after grouping, then the ratio of the reduced components is $(j-k)/j$.

\subsubsection{Baseline}

Xiaoyi et al.~\cite{zhang2021screen} propose a similar approach to group UI components for efficient navigation.
They develop multiple heuristics that group UI components based on their UI types, sizes, and spatial relationships on the rendered phone screen.
Considering similar application scenarios, we choose their method as the baseline for our experiments.

\begin{table}[]
\setlength{\abovecaptionskip}{0pt}
\setlength{\belowcaptionskip}{0pt}
  \caption{Evaluation Results of Accuracy of GUI Grouping}
  \label{tab:groupAcc}
\begin{tabular}{|l|l|l|l|ll|ll|}
\hline
\multirow{2}{*}{ID} &
  \multirow{2}{*}{App} &
  \multirow{2}{*}{Category} &
  \multirow{2}{*}{\#Installation (Million)} &
  \multicolumn{2}{c|}{ExactMatch} &
  \multicolumn{2}{c|}{ReducedUI} \\ 
 &             &                   &     & Ours & BL~\cite{zhang2021screen} & Ours & BL~\cite{zhang2021screen} \\ 

\midrule

1                    & iQIYI Video & Entertainment     & 5   & 0.86 & 0.71   & 0.62 & 0.49   \\
2                    & Coursera    & Educational       & 10  & 0.72 & 0.61   & 0.55 & 0.37   \\
3                    & Evernote    & Productivity      & 100 & 0.80 & 0.77   & 0.61 & 0.43   \\
4                    & Kodi        & Tool              & 50  & 0.85 & 0.81   & 0.62 & 0.52   \\
5                    & Pinterest   & Lifestyle         & 500 & 0.74 & 0.62   & 0.56 & 0.40   \\
6                    & Wonder      & Art \& Design     & 5   & 0.79 & 0.72   & 0.53 & 0.47   \\
7                    & Fiverr      & Bussiness         & 10  & 0.83 & 0.69   & 0.55 & 0.51   \\
8                    & ABC listen  & Music \& Audio    & 1   & 0.91 & 0.81   & 0.64 & 0.56   \\
9                    & Fitbit      & Health \& Fitness & 50  & 0.84 & 0.78   & 0.56 & 0.59   \\
10                   & Kik         & Communication     & 100 & 0.76 & 0.73   & 0.52 & 0.45   \\ 

\midrule

\multicolumn{4}{|c|}{Average}                                  & \textbf{0.81} & 0.73   & \textbf{0.58} & 0.48  \\

\bottomrule

\end{tabular}
\end{table}

\subsubsection{Results}
Table~\ref{tab:groupAcc} summarizes the information of the selected 10 apps and the accuracy results of the GUI grouping.
The column \emph{\#Installation} demonstrates the number of app installations. 
The Fleiss Kappa value of the first grouping results for three evaluators of our 10 apps are all between 0.91 and 1. Three evaluators discuss the different parts and finally agree on a unified final result. 
The subcolumn \emph{Ours} and \emph{BL} show our and the baseline' experimental results in the exact match rate and reduced UI components rates, respectively. 
Our approach achieves 0.81 in average exact match, which is 10.96\% higher than the baseline \emph{BL}~\cite{zhang2021screen} (0.73).
Our approach reduces isolated GUI components in GUI pages by an average of 58\%, which is 20.83\% higher than the baseline (48\%).
Both results demonstrate the effectiveness of our GUI grouping algorithms.

Figure~\ref{fig:exampleGroup} depicts the results of a pair of our method and baseline's method after grouping, with the first subplot representing our method and the second representing the baseline's.
Our approach splits the GUI page into 5 GUI groups and other GUI components. 
Correspondingly, the baseline separates the page into 8 distinct groupings and components.
Clearly, our grouping results are more precise and overlook fewer individual components. 
For example, in our group 2 and the corresponding groups 2 and 3 of the baseline, our method considers the possible related information surrounding the text, successfully groups the related images on the right side together, and merges the adjacent groups with the same structure, whereas the grouping result of the baseline method omits the related image data. 
In more complex scenarios, such as group 3 and group 4 in subfigure (a), corresponding to groups 4, 5, 6, and 7 in subfigure (b), the results of groups 4 and 5 in subfigure (b) omit the majority of the information since baseline's approach cannot handle the case of numerous lines of text. 
Our approach is built with more general atomic, row, and multi-row grouping methods, so that the group results contain as much important data as feasible, and reduces the number of groups on a GUI page to expedite the subsequent conversion operation.

\begin{figure}[htbp]
\centering
\begin{minipage}[t]{0.48\textwidth}
\centering
\includegraphics[width=6cm]{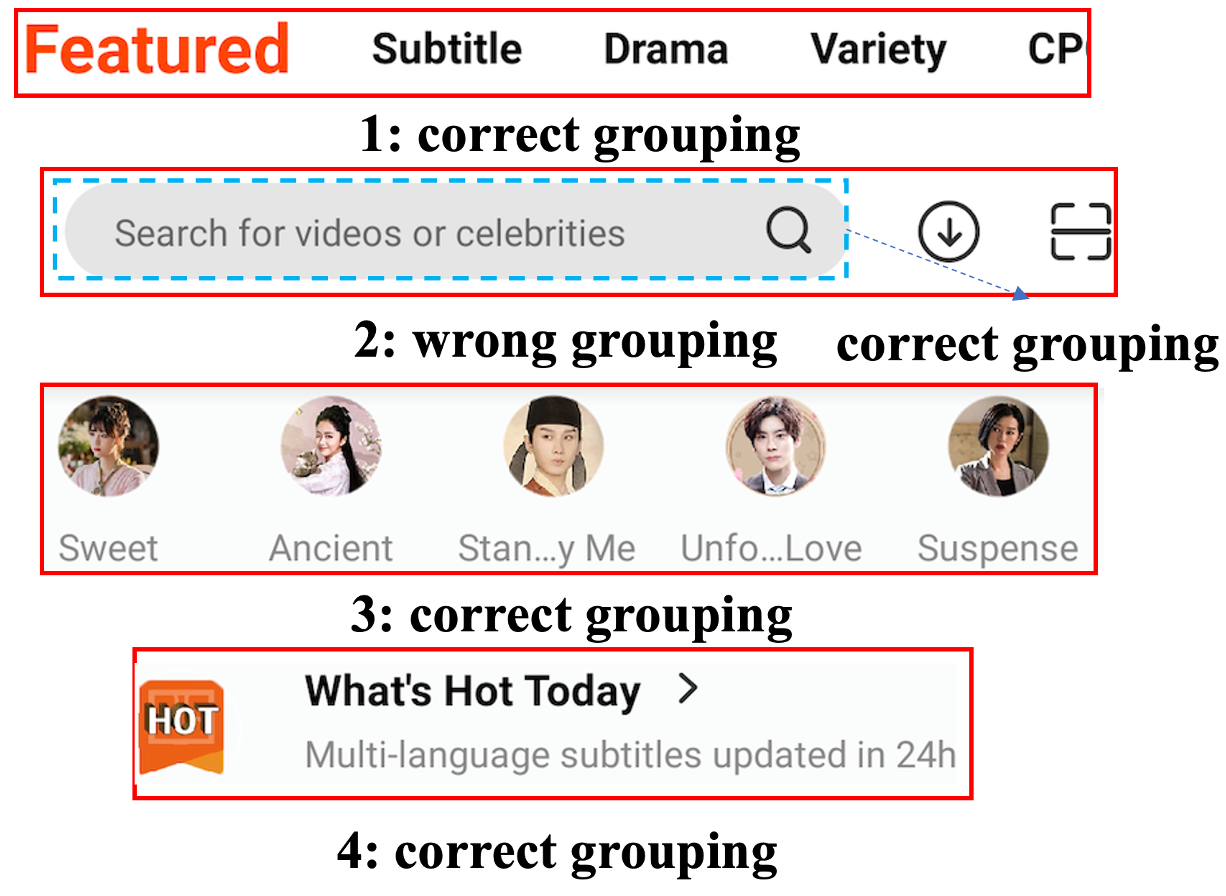}
\caption{Examples of Correct/Wrong GUI Grouping}
    \label{fig:exampleGroup}
\end{minipage}
\begin{minipage}[t]{0.48\textwidth}
\centering
\includegraphics[width=6cm]{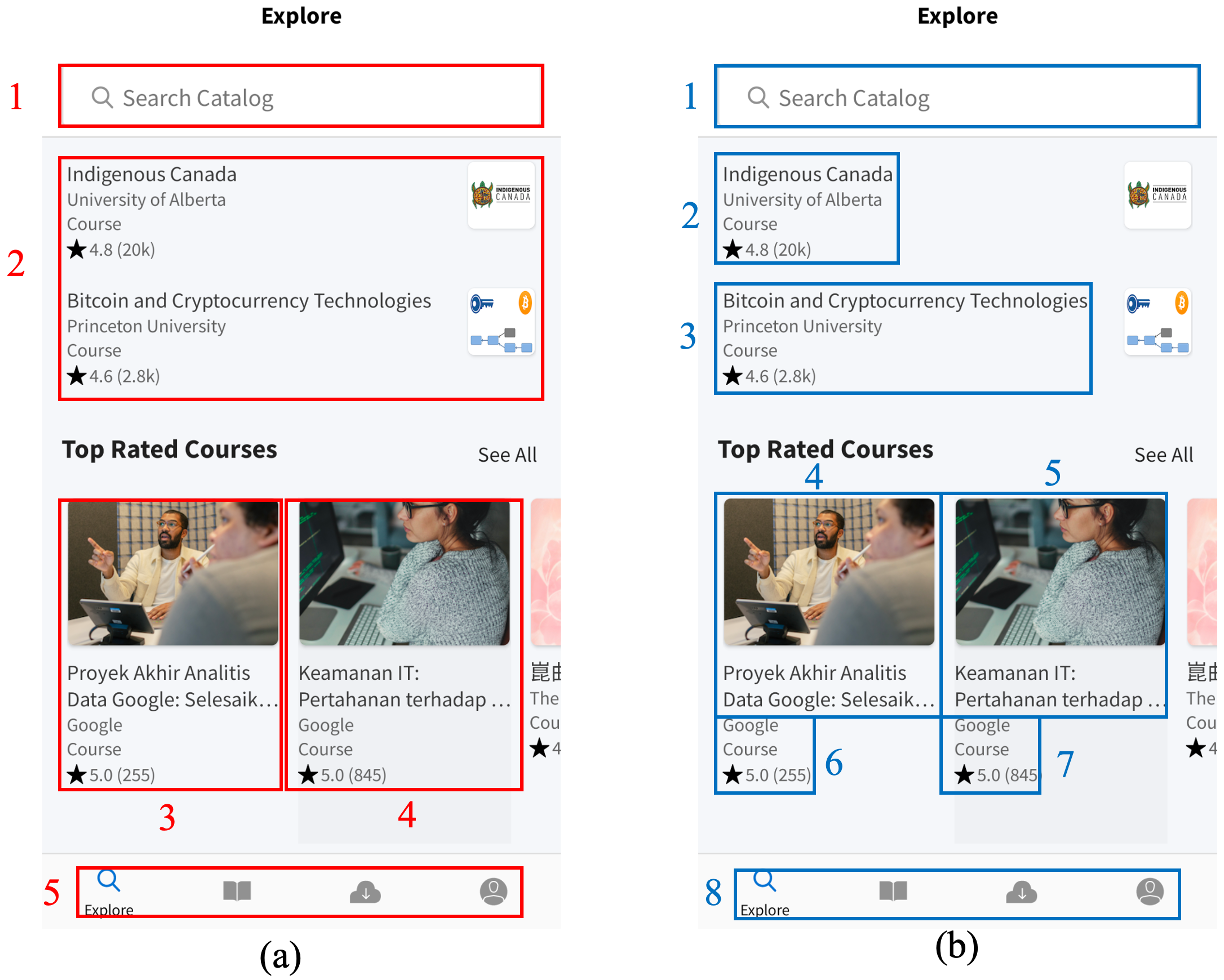}
\caption{Examples of Our and Baseline's Grouping Results. The first one is our grouping results and the second one is baseline's grouping results.}
 \label{fig:exampleGroup}
\end{minipage}
\end{figure}
\setlength{\textfloatsep}{10pt}

\subsection{Accuracy of GUI Conversion}
Once the TV GUI DSLs have been generated, we convert them into source code and then run them to get the rendered run-time GUI pages.
On the one hand, the impact of the same run-time GUI can be accomplished through a variety of code. 
On the other hand, there is a gap between the GUI source code and the rendered GUI effect, and the GUI source code does not reflect the rendered GUI effect in its entirety~\cite{rountev2014static}. 
To demonstrate the efficacy of the approach, we choose to evaluate the rendered effect of apps running through the translated DSLs.
We perform both an automatic evaluation and a user study to evaluate the performance of the whole automated GUI conversion approach.
All 589 pairs in Section~\ref{sec:rq2} are used in the automated evaluation to objectively evaluate the overall effect of the method.
The quality of visual transformation is strongly dependent on human perception, so we also include a user study to assess our approach's performance.
Due to the great efforts of the user study, we randomly sample 42 (7\%) GUI pairs among 589 pairs from 8 apps in 6 categories.

\subsubsection{Evaluation Metrics}


According to recent studies~\cite{chen2018ui, moran2018automated}, when using phones and TVs, users largely rely on the layout of images and texts to understand GUIs. 
Uniform GUIs facilitate user adaptability from mobile to TV app. Additionally, a consistent GUI promotes a reduction in developmental cost and time by enabling the reuse of code and design elements~\cite{mendoza2013mobile, lowdermilk2013user}.
Therefore, we substantiate the accuracy of GUI conversion by quantifying the similarity between TV GUIs that are automatically generated and those manually designed, serving as the ground truth in this study.
The mIOU~\cite{jaccard1912distribution}, which could effectively evaluate the layout gap between each type of the component in one GUI page, has been widely used in GUI evaluation~\cite{kumar2011bricolage,laine2021responsive,zheng2019content,chen2018ui,xie2020uied,zhao2021guigan,moran2018detecting}.
Based on its characteristics and suitability for our specific study, we select mIoU as the metric to evaluate the layout similarity of generated TV GUIs and ground truths.
The TV versions in phone-TV GUI pairs have been redesigned and optimized with a more logical layout of text and images in the GUI to accommodate the features of the TV.
So we select these redesigned TV GUIs as the ground truth of the corresponding phone GUI in the automated evaluation.

The mIoU (Mean Intersection Over Union), also known as the Jaccard Index, is a prominent image segmentation assessment measure that computes the IOU for each class before calculating the average over classes.
The IoU is calculated by dividing the overlap area between predicted class positions and ground truth by the area of union between predicted position and ground truth.
This is computed by:
\begin{equation}
    mIoU = \frac{1}{k}\sum_{i=0}^k\frac{TP(i)}{TP(i) + FP(i) + FN(i)}
\end{equation}
where $k$ means $k$ classes in both images, $TP(i)$, $FP(i)$ and $FN(i)$ represent the distribution of true positive, false positive, and false negative of $i_{th}$ class between two compared images. 

Our empirical study in section~\ref{sec:rq2} shows that the current GUIs of paired phones and TVs often do not have strict correspondence.
Besides, whether a GUI design makes sense depends significantly on the user's subjective perception. 
A GUI with a low mIOU in the automatic evaluation may be deemed acceptable by some users. 
For example, some GUI may have a more reasonable GUI design than the ground truth.
These reasonable GUI designs got low scores because their layouts are not in line with the ground truth.
To eliminate the bias, we adopted two metrics~\cite{zhao2021guigan}, structure rationality and overall satisfaction for participants in the user study to rate the quality of the generated TV GUI by considering the characteristics of the TV apps. 
These metrics were inspired by the web GUI evaluation~\cite{kumar2011bricolage, laine2021responsive, zheng2019content} and image evaluation~\cite{hore2010image, ponomarenko2008color}.
First, structure rationality is used to evaluate component layout rationality, which refers to the placement of components in the GUI as well as the reasoning behind their combination and sorting.
Second, overall satisfaction is to evaluate the overall design's pleasing qualities. 
For each metric, the participants will give a score ranging from 1 to 5, with 1 being the least satisfactoriness and 5 representing the highest satisfactoriness.

\subsubsection{Baselines}
Desktop mode~\cite{huaweiDesk, Dex} is widely used in various smartphones, such as Samsung, OnePlus, Huawei, and Oppo.
It allows users to connect an external display to an Android smartphone or tablet to make content easier to view, just like on a TV or computer.
The desktop mode is optimized for larger displays with resizable windows and a different layout for GUIs.
HDMI adapter or WiFi is required to use the desktop mode.
In this experiment, considering the current use range and maturity of computer mode, we first choose Huawei EMUI desktop mode as the baseline, which is one of the earliest phone models to support desktop mode.

Currently, Google provides the big-screen responsive layout component for Android system~\cite{AndroidTV, AndroidTVApp, AndroidTVDe, adaptiveUI, googleTV}. 
When an Android app runs on different-sized screens, adaptive and responsive Android GUI components will adjust their positions and sizes to fit the screen size of each device.
In our empirical study, we found that part of the apps using these technologies automatically adapts to different screens.
Even though these technologies often offer a worse user experience than a fully hand-optimized GUI, it is nevertheless worthwhile to compare them to semi-automatic methods.
Thus,  our second baseline is the result of directly mapping adaptive Android GUI from phone to TV.
Unlike the desktop mode, no external equipment is required for direct mapping.

In the user study, we also compare our method to the redesigned TV GUI, which serves as the ground truth for the automated evaluation.
The comparison with the ground truth will provide a clear image of the efficacy gap between our method and the redesigned GUI, which requires extra work.

\subsubsection{Procedures}
In the automatic evaluation, we select redesigned TV GUIs in 589 pairs as our ground truths.
We use our proposed approach to convert every phone GUI to DSLs of TV GUI.
Then we use the client app on TV to generate code and render GUIs on the Android TV emulators~\cite{androidStudio}.
The emulator is configured with 4 CPU processors, 4 GiB of RAM, and 1 GiB of SD card. 
The API level version of the system image is 26.
When GUIs are rendered, we get the metadata of GUIs to generate their corresponding wireframes.
The content of pictures and text in the GUI is not taken into consideration since we are comparing the structure of the rendered GUI, not the pixels of the produced GUI, therefore we convert images to red blocks and texts to green blocks in wireframes.
In the same way, we generate wireframes of baselines and our approach's generated TV GUI.
Finally, we evaluate the mIoU between wireframes of our generated TV GUIs and ground truths.

In the user study, we recruit 20 participants who are professional designers and developers with more than 3-year Android development experiment for this user study.
We first introduce them with a detailed explanation of tasks and the two GUI evaluation metrics structure rationality and overall satisfaction. 
Meanwhile, we provide participants with all generated GUI designs from different methods and ask them to give the score of each GUI design in two metrics of the user study. 
Note that they do not know which TV GUI is from which method and all of them will evaluate the TV GUI design individually without any discussion.
For each test case, participants are asked to choose one GUI they think works best.

\subsubsection{Results}


Table~\ref{tab:whole} demonstrates average results on the testing set of our approach and three baselines.
The ground truth GUI, which requires extra engineering effort to redesign one-to-one, is undoubtedly more favorable than the other three automatically generated approaches.
In 19 cases out of 42, \emph{Ground Truth} is deemed the most efficient, followed by our method (15), \emph{Desktop mode (7)}, and Direct mapping (1).
It reaches the highest structure rationality (4.27) and overall satisfaction (4.20).
However, this approach requires customization for each GUI page, adding significantly to the engineering expenses and making it non-generalizable. 
Compared to ground truth's approach, ours performs marginally worse (0.17 lower on \emph{Structure Rationality} and 1.2 lower on Overall Satisfaction). 
Our approach produces the best results across 15 cases, which is quite similar to the ground truth (19), and we do not require the additional engineering costs associated with tailoring each GUI page.
Our approach outperforms the other two baselines in mIoU, overall satisfaction, and structure rationality by significant margins in both metrics.
Compared with the baseline \emph{desktop mode}, our approach achieves 21.05\%, 10.42\%, and 21.31\% improvement in mIoU, overall satisfaction, and structure rationality.
According to the experimental findings, our method outperforms the other two automated methods (\emph{Direct mapping} and \emph{Desktop mode}). 
Our approach can also produce comparable outcomes when compared to the ground truth without incurring additional expenses.

\begin{table}
  \caption{Average score of evaluation. * denote the statistical significance $p-value < 0.01$ with other two metrics.}
  \label{tab:whole}
  \scalebox{0.9}{
\begin{tabular}{|c|c|c|c|c|}
    \toprule
Approach &
   mIoU &
  \begin{tabular}[c]{@{}l@{}}Structure \\ Rationality\end{tabular} &
  \begin{tabular}[c]{@{}l@{}}Overall \\ Satisfaction\end{tabular} & \begin{tabular}[c]{@{}l@{}}Vote \\ Best\end{tabular}\\
\midrule
Direct mapping & 0.07 &  2.44          & 2.67         & 1\\
Desktop mode   & 0.19 & 3.38          & 3.55           & 7\\
Ground truth & 1 & \textbf{4.27}          & \textbf{4.20}           & \textbf{19} \\ 
Ours           & \textbf{0.23} & 4.10* & 3.92*   & 15\\
\bottomrule
\end{tabular}
}
\end{table}

\begin{figure}
    \centering
    \includegraphics[width = 0.5\linewidth]{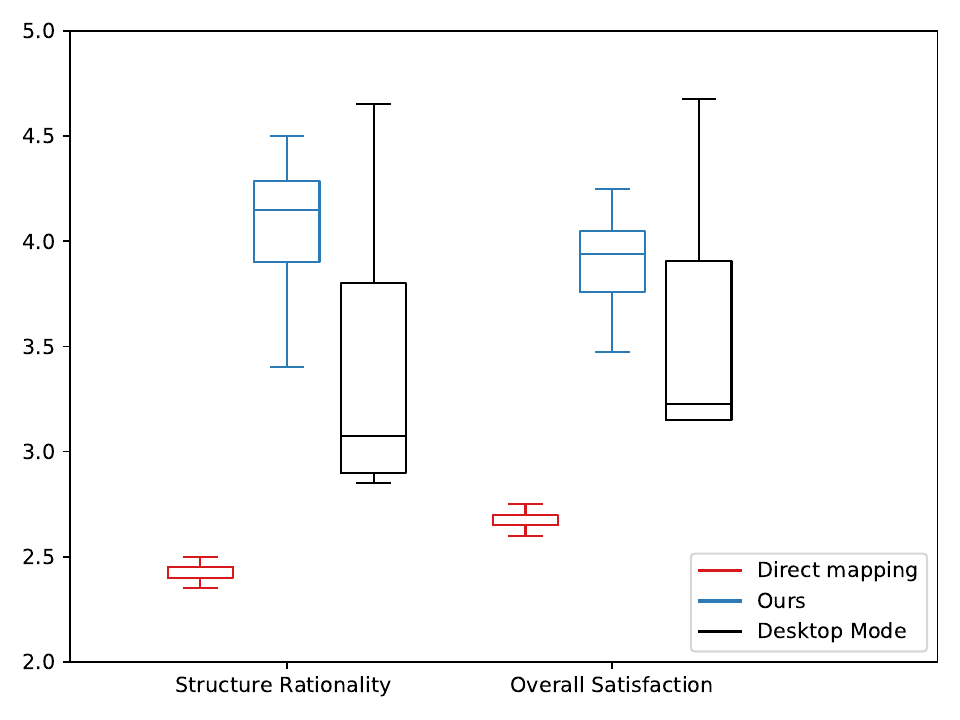}
    \caption{Score distribution of the structure rationality and overall satisfaction of three approaches}
    \label{fig:boxPlot}
\end{figure}

To further analyze our method and the other two automatic methods (\emph{Direct mapping} and \emph{Desktop mode}), we plot the boxplots of the scores of these three methods over \emph{Structure Rationality} and \emph{Overall Satisfaction} in user study in Figure~\ref{fig:boxPlot}.
In both box plots, the gap between our first and third quartiles was lower than in \emph{Desktop mode}, indicating that our ratings are more concentrated and stable.
The maximums of \emph{Desktop mode} in both box plots are higher than ours.
There are a few cases that have been individually optimized for \emph{Desktop mode} display, so the scores are particularly high, which increases the maximum score of the whole. 
However, if the page is not individually optimized, the effect of \emph{Desktop mode} is significantly lower than that of our method.
The scores of \emph{Direct mapping} are all significantly lower than the other two, indicating that users generally do not accept this conversion.

To understand the significance of the differences between the user study results of baselines and our approach, we carry out the Mann-Whitney U test~\cite{fay2010wilcoxon} on the overall satisfaction and structure rationality with \emph{Direct mapping} and \emph{Desktop mode}, respectively.
Since there are two baselines, we carry out the test between our approach with each baseline separately.
The test results in Table~\ref{tab:whole} show that our tool can significantly create better GUI conversion in overall satisfaction and structural rationality from phone to TV than \emph{Desktop mode} (both $p-value < 0.01$) and \emph{Direct mapping} (both $p-value < 0.01$).

\begin{figure*}
    \centering
    \includegraphics[width = 0.9\linewidth]{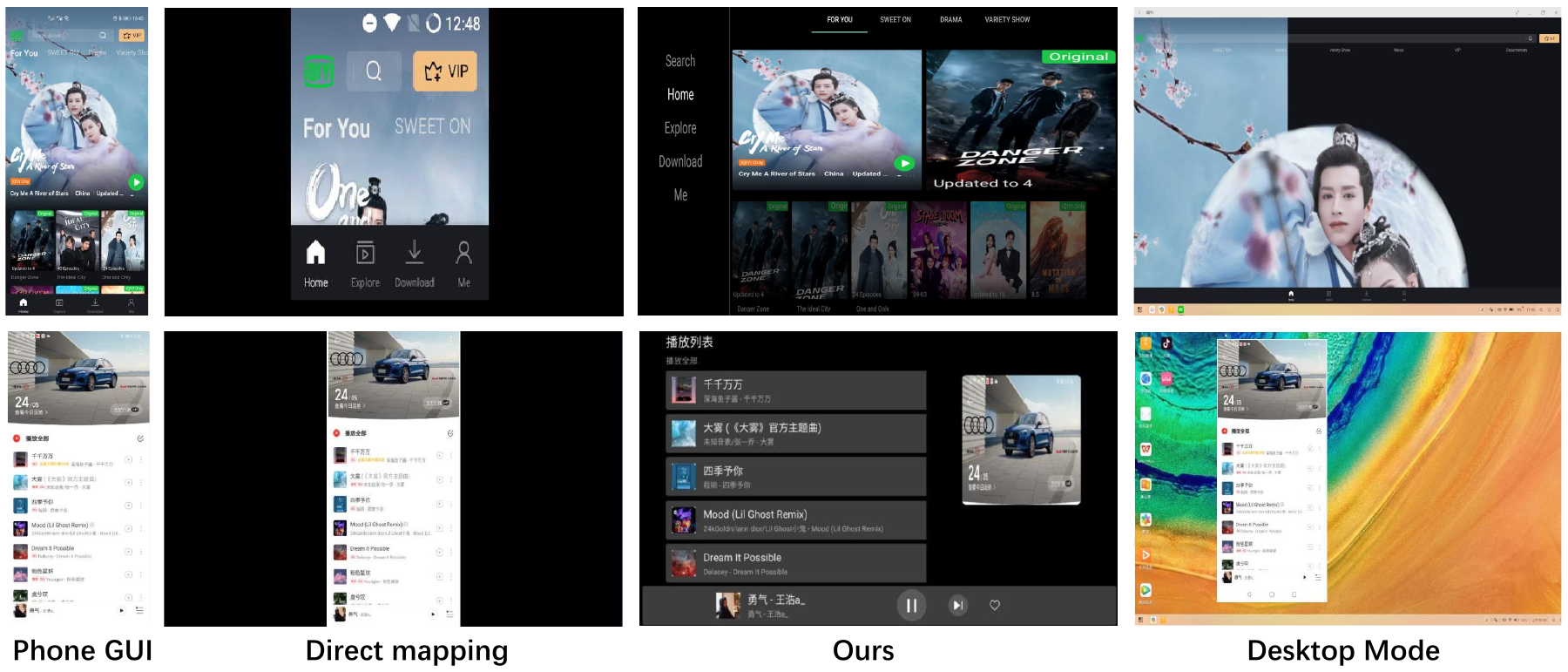}
    \caption{Conversion Examples by Direct Mapping, our tool, and Desktop Mode}
    \label{fig:examples}
\end{figure*}

To better illustrate our results, Figure~\ref{fig:examples} lists examples of generated TV GUI by different approaches.
For the case above in Figure~\ref{fig:examples}, the average structure rationality for \emph{Direct mapping}, our tool and \emph{Desktop mode} are 1.8, 4.4, and 3.2.
The average overall satisfaction for \emph{Direct mapping}, our tool and \emph{Desktop mode} are 1.7, 4.0, and 3.6.
For the case below in Figure~\ref{fig:examples}, the average structure rationality for \emph{Direct mapping}, our tool and \emph{Desktop mode} are 1.4, 4.6, and 1.5.
The average overall satisfaction for \emph{Direct mapping}, our tool and \emph{Desktop mode} are 2.0, 4.3, and 1.9.
Generally, for \emph{Desktop model}, only a few specific apps and pages are manually optimized, so when testing with various types of pages, uncustomized pages are original phone pages without optimization, resulting in low scores.
Faced with large TV screens, most cases are very blunt without considering the design criteria of the current platform's widgets and users' usage habits.
\emph{Direct mapping} places all components as where they are on the phone, resulting in a very poor user experience for TV users.
Besides, it can only convert part of GUI components, which has poor universality.
This it is generally not accepted by users.

\begin{figure}
    \centering
    \includegraphics[width = 0.7\linewidth]{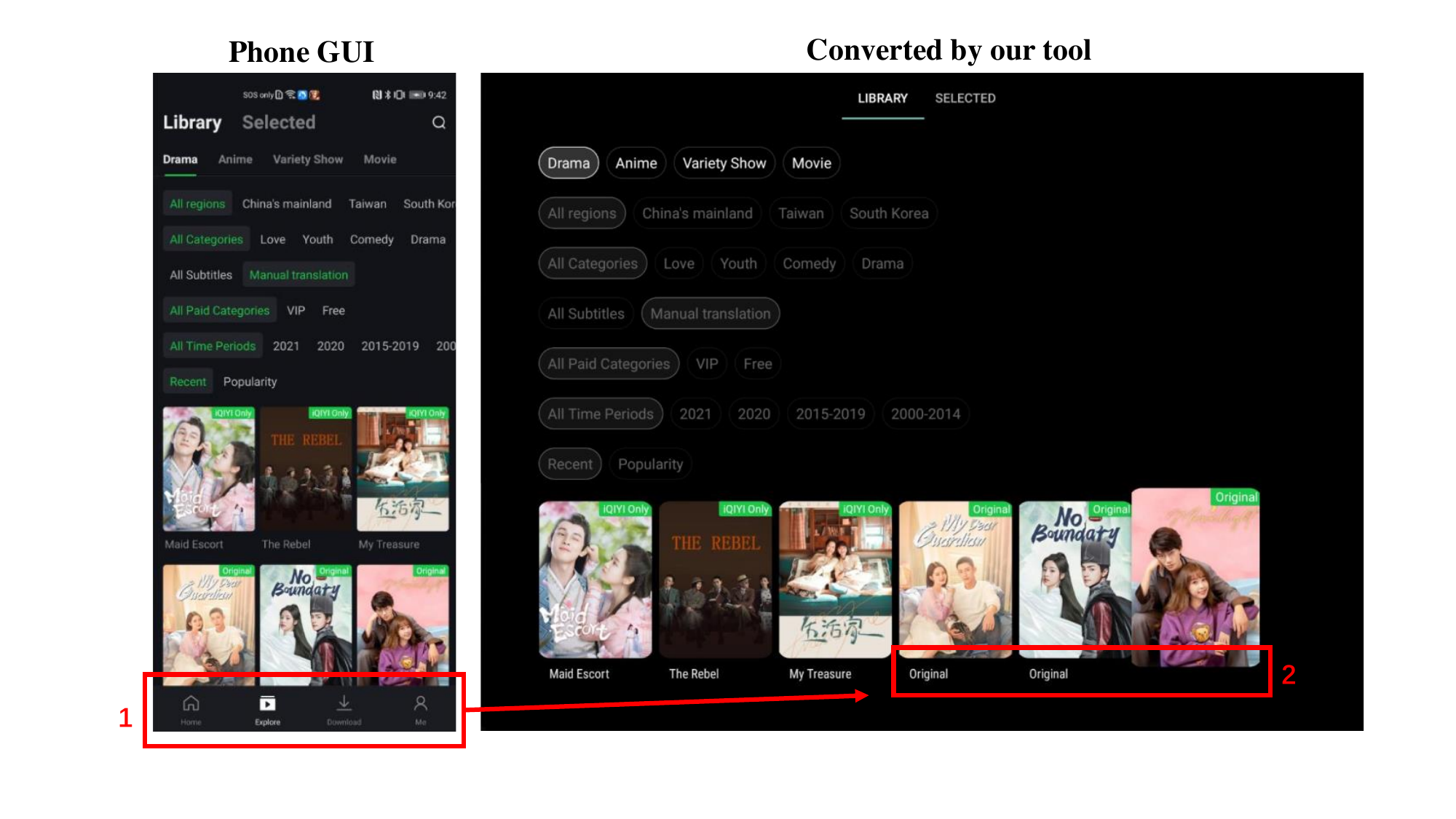}
    \caption{An example of metadata data error}
    \label{fig:error}
\end{figure}

Our approach sometimes may also generate inappropriate TV mapping, especially for those non-standard GUI inputs.
For the phone GUI in Figure~\ref{fig:error}, promotional images of films in the second row do not show completely due to the size of the mobile screen.
In this case, UI Automator incorrectly provides us with the name of these films with details in \textit{Bottom Tab Layout}.
Thus, our approach fails to get the correct name of films due to the limitation of UI Automator, just like the example shown in red boxes.
As our approach gains confusing hierarchies from UI Automator, it does not correctly identify the \textit{Bottom Tab Layout}.
Therefore, the approach fails \textit{Bottom Tab Layout} to convert it into a TV \textit{Channel}.

\section{Usefulness Evaluation}
According to the findings of our empirical study in Section~\ref{sec:supportTV}, more developers prefer to redevelop the GUI pages of new TV apps. 
Even while experienced developers can produce the matching TV GUI pages quickly, it still affects the development efficiency and wastes some valuable resources in the phone GUI.
We carry out a user study to evaluate the usefulness of our generated TV GUI for bootstrapping corresponding TV GUI design and implementation by real-world developers.

\subsection{Procedures}
We recruit 6 participants who are all working in software companies and have at least one-year Android GUI developing experience.
Participants are required to design and implement the corresponding conversion TV GUI by referring to the given Phone GUI.
We provide 6 phone GUIs which have covered main 12 phone GUI groups.
The official corresponding designed TV GUIs of 6 phone GUIs are also collected for the subsequent satisfaction evaluation. 
Participants are required to design and implement the layout skeleton and set component properties, including the view type, size, order, and padding.
Note that participants are allowed to replace some component types with placeholders without affecting the rendering and overall design.

The study consists of two groups of three participants: the experimental group, consisting of three participants, is asked to proceed on the basis of our tool, and the control group, which starts work from scratch on a TV app design.
The experiment group is allowed to use our tool to automatically generate a draft TV GUI and update the generated source code to re-design TV GUI directly.
Each member in the experimental group learns in advance how to use our tool to generate source code and render it on TV. 
Participants in both groups have comparable development
experience by pre-study their developing background to ensure both groups have similar expertise in total.
All participants are only allowed to use Android Studio and Java to avoid bias and have up to 20 minutes for each implementation.
The decision to allot 20 minutes for the experiment is made in close consultation with the participating developers. 
Three academic Ph.D. students who are not involved in the study are asked to evaluate 6 participants' results and rate their satisfaction on a five-point scale, with 5 being the highest and 1 the lowest.
The satisfaction metric\cite{kumar2011bricolage, laine2021responsive, zheng2019content} is to evaluate the overall pleasing qualities of this GUI page.
The evaluators must determine if the layout, content, and UI type of all UI components on this GUI page are appropriate.
When rating, raters do not know which TV GUI is developed by which group and the manual-designed TV GUI by companies are used for the reference.
Similar to the accuracy evaluation, we use the Fleiss Kappa value~\cite{fleiss1971measuring} to measure the agreement among the three raters.
Fleiss Kappa values in the range of [0.01, 0.20], (0.20, 0.40], (0.40, 0.60], (0.60, 0.80], and (0.8, 1] correspond to the slight, fair, moderate, substantial, and almost perfect agreement, respectively.
If the Fleiss Kappa value is less than 0.8, the divergent cases will be discussed, analyzed, and re-scored by the three raters until the Fleiss Kappa value is greater than 0.8.

We provide 6 participants with the same development environment and resources and record the time it takes participants to complete each TV GUI.
In this study, the primary metric centers around the discernible outcomes between the control and experimental groups. A notable difference between these groups serves as evidence of our tool's effectiveness in enhancing the development process.


\begin{figure}
\begin{minipage}{0.5\linewidth}
    \centering
    \includegraphics[width=\linewidth, height = 4 cm]{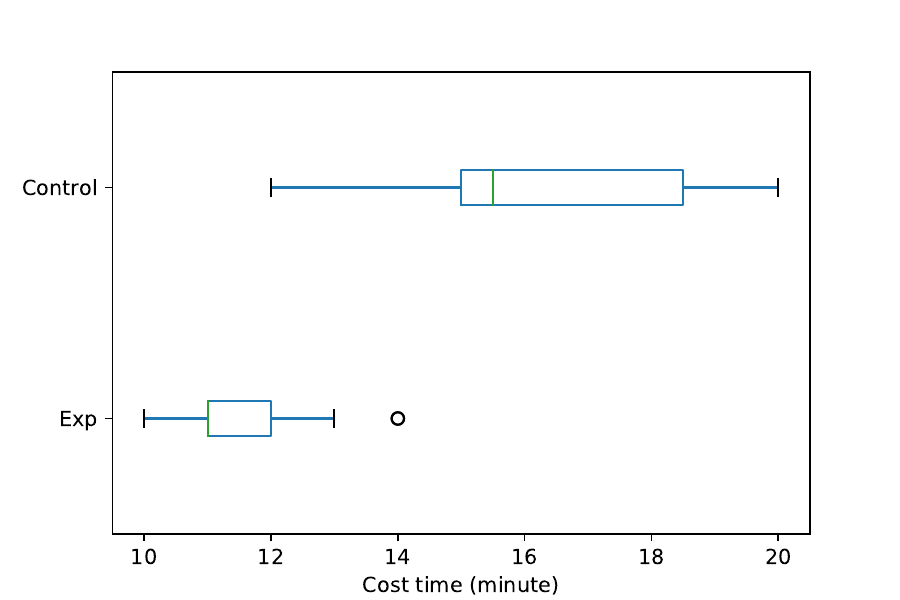}
    \caption{Distribution of cost time and satisfaction scores of control and experimental groups}
    \label{fig:boxUsefulness}
\end{minipage}
\noindent
\begin{minipage}{0.45\linewidth}
\captionof{table}{Average cost time and satisfaction scores of control and experimental groups. ** denote the statistical significance $p-value < 0.01$.}
\begin{tabular}{c|c|c}
\hline
  Indicator  &  Control  &  Experiment \\
   Cost Time (min)  & 16.28 & 11.56** \\
  Average Score & 1.66 & 3.35** \\
\hline
\end{tabular}
\label{tab:aveUsefulness}
\end{minipage}
\end{figure}

\subsection{Results}
The box plot in Figure~\ref{fig:boxPlot} and the average score in Table~\ref{tab:aveUsefulness} show that the experimental group implements TV GUI conversion faster (average of 11.56 minutes) with a higher satisfaction score (average of 3.35) than the control group (average of 16.28 minutes and 1.66 satisfaction score).
Members of the experimental group use the DSL to design the GUI, but members of the control group write actual code to design the GUI.
Therefore, their working implementations are different, and the outcomes of the control group members are closer to the actual app implementation.
Theoretically, different working implementations could lead to bias in the results, but all invited professionals have at least one year of Android development experience. 
They master the Android GUI development procedure. 
Some developers use placeholders to replace some UI components, allowing them to focus on developing the GUI rather than completing the code logic and syntax. 
While the working implementations of the two groups may not be equal, the efforts of redesigning and constructing an adequate TV GUI are the decisive factor in determining the outcomes of the two groups experiments.
Our tool successfully and effectively assists developers in developing more suitable TV GUIs faster, taking into account the apparent time difference between 11.56 minutes and 15.6 minutes on average.
In experiments, two participants in the control group fail to finish at least one GUI conversion within 20 minutes.
However, everyone in the experimental group completed the conversion within 20 minutes and had time to personalize the GUIs. This results in significantly higher satisfaction scores than the control group (an average score increase of 1.64).

\begin{figure}
    \centering
    \includegraphics[width = 0.7\linewidth]{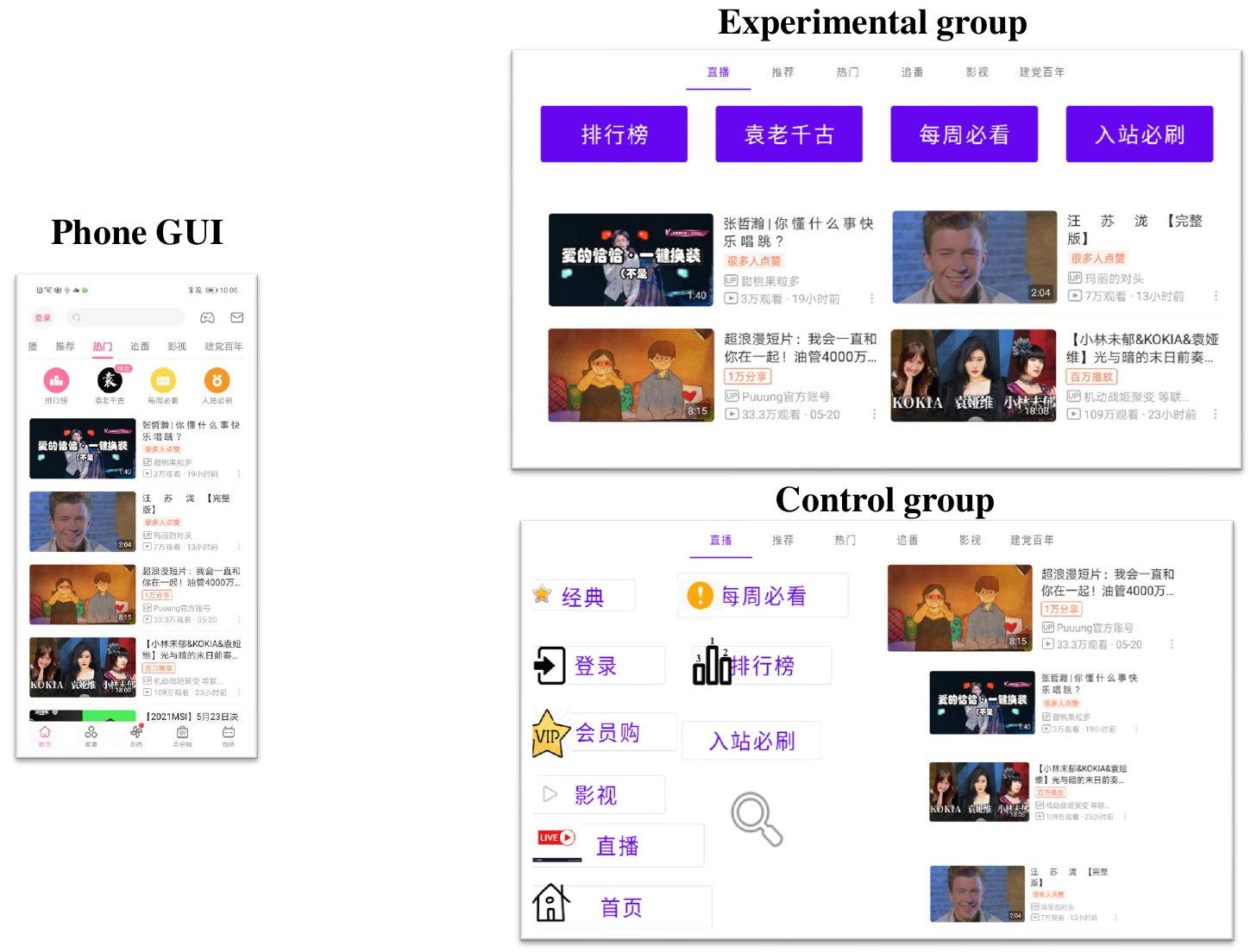}
    \caption{An example of experimental and control group}
    \label{fig:control}
\end{figure}

Figure~\ref{fig:control} shows an example of two TV GUI conversions from the experimental and control group.
Note that we only evaluate the structure and overall soundness of the GUI, so we do not evaluate the content in the participant's GUI, allowing placeholders.
When designing GUI, members in the experimental group can refer to the output layout of our tool, and can customize the TV GUI on this basis.
Our tool is an excellent inspiration for the experimental group, saving them time in designing and developing the TV GUI and giving them a bottom line for their work.
We also carry the Mann-Whitney U test on the cost time and satisfactory scores. 
The test results suggest that our tool can significantly help the experimental group convert phone GUI faster($p-value < 0.01$) and create better TV GUI ($p-value < 0.01$).

A participant in the experimental group commented on our tool that ``\textit{The automatic conversion results provided by the tool can give me good development design guidance and hints, greatly improving my development efficiency}''.
Overall, the user study results provide preliminary evidence of the usefulness of our tool.
The majority of participants think that the GUI DSL we built is simple to comprehend and use.
They said that our GUI DSL is a useful solution to address the present issue of GUI code reuse caused by the version differences between mobile phone systems and television systems.
Our GUI DSL also makes it simple for designers without any programming background to collaborate on projects.
Participants also pointed out some flaws in our existing approach. 
It is not appropriate for converting real-time demanding apps like games since the generated GUI needs to be re-rendered by the client app on the TV.
Because the image on the mobile GUI isn't designed with a large-screen TV, sometimes the image will seem stretched on the TV GUI after conversion, which can negatively impact the user experience.
To address these issues, we will design more efficient GUI rendering technology solutions and design the display of image resources after screen size conversion in future work.

\section{Threats to validity}
Potential threats to validity in our user study for performance evaluation primarily stem from subjectivity or bias inherent in participants' abilities and backgrounds. 
Similarly, the experiment on GUI grouping accuracy could be influenced by individual interpretations of `logically related UI components.' We have taken measures to mitigate these threats: participants are all GUI development professionals with diverse skills and experiences. We've provided comprehensive examples of 'logically related' components pre-experiment to ensure a shared understanding. We refrain from revealing which results are ours or the baselines' during the user study. Moreover, results are objectively assessed by comparing the mIoU to the redesigned native TV GUI, with tool performance evaluated via a blend of user study and automated evaluation results.

In the usefulness evaluation, we primarily relied on `satisfaction score' and implementation time as the evaluative metrics for our approach and the baselines. This may not fully encapsulate all elements pertinent to GUI usability and aesthetics. 
However, while the `satisfaction score' is central to our usefulness evaluation, we employ other metrics in separate experiments designed to measure different aspects of our tool. Collectively, these various metrics corroborate the effectiveness of our proposed tool, contributing to a comprehensive assessment of its performance.
To broaden the scope of the tool's efficacy evaluation, we intend to introduce additional valid metrics in future work.
The diversity in development environments and backgrounds of developers across various regions and organizations poses a potential threat to our study's validity, as we may unintentionally miss certain perspectives. To address this, we actively sought participants from varied genders, nationalities, and affiliations. In future research phases, we plan to augment our participant base, especially targeting underrepresented sectors and regions, to strengthen the generalizability of our findings.

In our conversion pipeline, challenges arise from accurately classifying certain GUI components and potential deviations from Android design principles, leading to subsequent mapping errors. Additionally, current tools like UI Automator may not effectively capture metadata for specific third-party GUIs and $WebView$ views. To address these internal validity concerns, we default to a Grid Layout for unclassified GUI groups, ensuring information consistency, as described in Table~\ref{tab:matchRules}. Externally, Android's device fragmentation poses threats. To mitigate this, we utilize a universal GUI DSL for code synthesis, with a DSL translator tailoring the conversion to each TV's specifications, while leveraging predefined and quality-assured TV GUI libraries to enhance conversion reliability.

A potential limitation in the generalization of our UI grouping and conversion based on heuristic rules presents a threat to validity. Though our method is designed for specific GUI paradigms, empirical studies across varied GUI designs attest to its efficacy in a considerable portion of the app ecosystem. In Section~\ref{sec: appCollection}, we outline our data collection from Google Play and Dangbei, resulting in 2,805 TV apps. Our subsequent analysis, detailed in Section~\ref{sec:groupAlignment}, covered 1,405 TV-phone app pairs and yielded favorable results in our manual verification of 589 real-world GUI pairs. Despite its specificity, our method's effectiveness is evident in a significant app ecosystem segment. Google's Android TV apps, emphasizing card-like layouts, align with our grid layout template, ensuring adaptability with OR constraints. Looking ahead, our framework, optimized for mobile-to-TV conversions, serves as a robust foundation for extending into areas like mobile-to-smartwatch or augmented reality, as GUI paradigms evolve. Future research could focus on our method's scalability and potential integration of deep learning for enhanced UI grouping and component synchronization.

\section{Related Work}

\subsection{GUI implementation}
Implementing a GUI focuses on making the GUI work with proper layouts and widgets of a GUI framework.
Nguyen and Csallner~\cite{nguyen2015reverse} reverse engineer the UI screenshots by rule-based image processing method and generate GUI code. 
They support only a small set of most commonly used GUI components. 
More powerful deep learning-based methods~\cite{beltramelli2018pix2code, chen2018ui, moran2018machine, hu2019code} have been recently proposed to leverage the big data of automatically collected UI screenshots and corresponding code.
Some recent works explore issues between UI designs and their implementations. 
Moran et al.~\cite{moran2018automated} check if the implemented GUI violates the original UI design by comparing the image's similarity with computer vision techniques. 
The follow-up work~\cite{moran2018detecting} further detects and summarizes GUI changes in evolving mobile apps. 
Similarly, the semantic vector for the UI design from our work can help detect the inconsistency among UI designs within the same app.

Different from those works on GUI implementation which are highly related to conventional GUI development, we are targeting at specifically GUI projecting from small-screen mobile phones to the corresponding one on large-screen TVs.

\subsection{GUI component grouping}
There are some similar components of clustering and page segmentation algorithms in web page analysis.
Yandrapally et al.~\cite{yandrapally2020near} propose a near-duplication detection algorithm to study near-duplication components in web pages.
They characterize and merge functional near-duplicates by summarizing categories in existing web pages.
Crescenzi et al.~\cite{crescenzi2005clustering} propose a structural abstraction clustering algorithm for web pages that groups web pages based on this abstraction.
To assess end-to-end web tests, Yandrapally et al.~\cite{yandrapally2021mutation} utilize VIPS~\cite{cai2003vips} for web page segmentation and XPaths of web elements inside these fragments for establishing their equivalence.
VIPS~\cite{cai2003vips} is a popular page segmentation, it proposes an automatic top-down, tag-tree independent approach to detect web content structure. 
Mahajan et al.~\cite{mahajan2018automated, mahajan2021effective} design a clustering technique for web elements that are based on a combination of visual aspects and DOM-based XPath similarity.

Although our study focuses on the grouping and segmentation of mobile GUIs, their work on the clustering of web components enlightens us, and we incorporate their insights into mobile GUI grouping.

\subsection{GUI migration across platforms}
\label{sec:guiMigration}
Due to the difficulty of GUI migration across different platforms, very few related works are carried out to solve this problem.
Pihlajam~\cite{pihlajamaki2016desktop} observed multiple games across web and mobile platforms for summarizing several UI patterns for user interface adaptation. 
Wong et al.~\cite{wong2008transformation} developed a scalable GUI system that dynamically transforms platform-specific GUI widgets migrated within an application between any of a plurality of heterogeneous device platforms. 
Verhaeghe et al.~\cite{verhaeghe2019gui} developed a set of rules for GUI Migration using MDE from GWT to Angular 6.
Hu et al.~\cite{hu2023pairwise} curated a pairwise dataset encompassing phone-tablet GUI interfaces. Utilizing this dataset, they trained a generative model dedicated to facilitating automated GUI conversion between mobile phones and tablets.

Although these studies explore the mapping (or partial mapping) between different platforms and programming languages, none of these works are investigating app GUI adaptation between phones and TV. 
Instead of the white-box migration or layout migration, our study focuses more on black-box GUI adaptation without the source code of the original  GUI on the phone.
That black-box characteristics and significant differences in screen sizes motivate us to develop a brand new approach to bridge the gap between phone and TV GUI.
Besides, time and resource consumption must be considered because of phones' current hardware limitations.

\subsection{Practical tools in industry}
Finally, it is worth mentioning some related non-academic projects.
There are many third-party libraries or frameworks which support cross-platform adaptation such as React.js~\cite{React}, Flutter~\cite{Flutter}, and also default Android development ~\cite{AndroidTV}. 
Although these frameworks can cover multiple platforms such as smartphones, desktops, and tablets, TV is rarely covered due to its ultra-large screen. 
Developers have to commit much additional effort to design new layouts that can be easily understood from 10 feet away and provide navigation that works with just a directional pad and a select button to make their app successful on TV devices. 
That is one reason why few apps support GUI adaptation, and the small number of smart TV users further discourages app developers.

Samsung supports screen mirroring from the Samsung device to the TV including photos, videos, presentations, and games~\cite{Dex}. 
There are also other similar connections between smartphones and TV such as Chromecast\cite{chromecast}, Xiaomi TV stick~\cite{miTV} or wired connection via HDMI, etc.
Most of them are working well in only video projection or some customized apps, and they require additional support from the TV side.
Directly showing phone GUI on TV will bring many rendering issues such as small components, large black margins, unreasonable interaction, etc. 
To overcome those issues, we propose an automated approach to project the phone GUI to the TV on the run time.

\section{Conclusion}
At the moment, adaptive technologies between phone and TV GUI are unable to fulfill the demands of app developers, and the cost of developing and maintaining new applications is prohibitive, therefore building an automated GUI conversion tool from phone to TV is challenging but worthwhile work for developers.
An automated approach for converting phone GUI to TV GUI is presented in this study.
Before proposing our approach, we carry out empirical studies to explore how many current apps support TV displays and how current apps convert phone GUI to TV GUI.
Our tool consists of four integral stages: GUI components grouping, template matching, layout optimization, and DSL for code synthesis.
Finally, we convert the generated GUI DSL to source code for rendering the final TV GUIs.
Our approach offers obvious benefits over existing mainstream technologies, according to an automated evaluation in 589 valid phone-TV GUI pairs and a user study from 20 Android professionals.
Besides, a pilot user study also illustrates the usefulness of our tool for app developers.

In the future, we will keep improving our algorithm for generating mapping GUIs from phone to TV.
With more and more apps developed for supporting TV, we will construct a large parallel corpus of TV-phone GUI pairs.
Based on that data, we will develop an end-to-end machine learning algorithm that will be more generalized than the current approach.
On the other hand, we will extend our tool to other platforms such as smartwatches, tablets, and vehicle screens.

\section*{Acknowledgements}

Grundy is supported by ARC Laureate Fellowship FL190100035.


\bibliographystyle{ACM-Reference-Format}
\bibliography{reference}

\appendix

\end{document}